\documentclass[10pt]{article}

\usepackage{lmodern}
\usepackage[latin1]{inputenc} 
\usepackage[T1]{fontenc}
\usepackage[french,english,francais]{babel}
\usepackage{amsmath}
\usepackage{amsthm}
\usepackage{amssymb}
\usepackage{amsfonts}
\usepackage{amscd}
\usepackage{latexsym}
\usepackage{graphicx}
\usepackage{hyperref}
\usepackage{epsfig}
\usepackage{bm}
\usepackage{dsfont}
\usepackage{color}
\usepackage{stmaryrd}
\usepackage{etoolbox}
\patchcmd{\thebibliography}{\section*}{\section}{}{}

\makeatletter
\@addtoreset{equation}{section}
\makeatother
\newtheorem{theorem}{Theorem}[section]

\newtheorem{remark}{Remark}[section]
\newtheorem{proposition}{Proposition}[section]
\newtheorem{lemma}{Lemma}[section]
\newtheorem{corollary}{Corollary}[section]
\newtheorem{definition}{Definition}[section]
\def\br{\begin{remark}\rm\small}
\def\er{\end{remark}}
\def\bt{\begin{theorem}}
\def\et{\end{theorem}}
\def\bd{\begin{definition}}
\def\ed{\end{definition}}
\def\bp{\begin{proposition}}
\def\ep{\end{proposition}}
\def\bl{\begin{lemma}}
\def\el{\end{lemma}}
\def\bc{\begin{corollary}}
\def\ec{\end{corollary}}
\def\beaq{\begin{eqnarray}}
\def\eeaq{\end{eqnarray}}

\newcommand\encadremath[1]{\vbox{\hrule\hbox{\vrule\kern8pt
\vbox{\kern8pt \hbox{$\displaystyle #1$}\kern8pt}
\kern8pt\vrule}\hrule}}
\def\enca#1{\vbox{\hrule\hbox{
\vrule\kern8pt\vbox{\kern8pt \hbox{$\displaystyle #1$}
\kern8pt} \kern8pt\vrule}\hrule}}

\newcommand{\td}{\tilde}

\newcommand{\eq}[1]{eq.~(\ref{#1})}

\newcommand{\Res}{\mathop{\,\rm Res\,}}

\newcommand{\beq}{\begin{equation}}
\newcommand{\eeq}{\end{equation}}
\newcommand{\beqq}{\begin{equation*}}
\newcommand{\eeqq}{\end{equation*}}
\newcommand{\bea}{\begin{eqnarray}}
\newcommand{\eea}{\end{eqnarray}}
\newcommand{\beaa}{\begin{eqnarray*}}
\newcommand{\eeaa}{\end{eqnarray*}}

\newcommand{\Tr}{\operatorname{Tr}}

\newcommand{\om}{\omega}

\begin{document}
\selectlanguage{english}
\begin{center}
\vspace{1cm}

{\Large \bf {Isomonodromic deformations of a rational differential system and reconstruction with the topological recursion: the $\mathfrak{sl}_2$ case}}

\vspace{0.5cm}

{Olivier Marchal}$^\dagger$\footnote{olivier.marchal@univ-st-etienne.fr},
{Nicolas Orantin}\footnote{nicolas.orantin@gmail.com},

\vspace{5mm}
$^1$\, Univ Lyon, Universit\'{e} Jean Monnet Saint-\'{E}tienne, CNRS UMR 5208, Institut Camille Jordan, F-42023 Saint-Etienne, France
\vspace{5mm}

\end{center}
\vspace{1cm}

{\bf Abstract :}
In this paper, we show that it is always possible to deform a differential equation $\partial_x \Psi(x) = L(x) \Psi(x)$ with $L(x) \in \mathfrak{sl}_2(\mathbb{C})(x)$ by introducing a small formal parameter $\hbar$ in such a way that it satisfies the Topological Type properties of Berg\`ere, Borot and Eynard. This is obtained by including the former differential equation in an isomonodromic system and using some homogeneity conditions to introduce $\hbar$. The topological recursion is then proved to provide a formal series expansion of the corresponding tau-function whose coefficients can thus be expressed in terms of intersections of tautological classes in the Deligne-Mumford compactification of the moduli space of surfaces. We present a few examples including any Fuchsian system of  $\mathfrak{sl}_2(\mathbb{C})(x)$ as well as some elements of Painlev\'e hierarchies.


\tableofcontents


\section{Introduction and summary of results}

The theory of isomonodromic systems has attracted a renewed attention in the recent years through its interaction with different topics in mathematics and physics such as Conformal Field Theories (CFTs) \cite{Harnad-KZ,Teschner-and-co,Reshetikhin}, random matrix models \cite{BM}, $\mathcal{N}=2$ SUSY theories \cite{CV}, Hitchin integrable systems and moduli spaces of flat connections on a Riemann surface \cite{Boalch,Teschner} or Nakajima quiver varieties \cite{Hiroe}. In particular, according to the isomonodromic-CFT correspondence, the tau function of an isomonodromic system should be equal to a conformal block of the corresponding theory, i.e. a section of the associated Verlinde bundle seen as a wave function obtained by geometric quantization of a moduli space of connections on the Riemann surface considered. From a physics point of view, this correspondence states that the tau function is a building block for correlation functions of the associated CFT.

On the other hand, the Topological Recursion (TR) \cite{EO}, a formalism originally developed for solving random matrix models \cite{CEO}, proved to be another  useful tool for studying different problems in mathematics in a unified way, allowing to bring together different fields of mathematics and physics \cite{Borot-review}. After it was proved that the topological recursion provides an efficient way of computing Gromov-Witten invariants of manifolds with semi-simple cohomology \cite{DBOSS,NS}, one of the most active and mysterious field of applications lies in the theory of  ``quantum curves'' where the recursion is expected to give a new way to compute the WKB expansion of a solution $\Psi(x,\hbar)$ of a given differential equation with respect to a formal parameter $\hbar$ \cite{QuantumCurveDModule,Quantum,ListSpecCurve,IS,ReviewNorbury,QuantumCurves}. In particular, if this differential equation comes from the quantization of the $A$-polynomial of a given knot, the formal series expansion of this solution is conjectured to provide a generating series of some associated knot invariants \cite{BEKnot}. The latter wave function can also be thought of as obtained by geometric quantization of some moduli space of connections on the boundary of a tubular neighborhood of the knot. 
On the other hand, when the differential equation defined by the quantum curve belongs to some isomonodromic systems, the partition function $Z = \exp \left(\underset{g=0}{\overset{\infty}{\sum}}\hbar^{2g-2} F_g\right)$ is equal to an associated isomonodromic tau function.

However, not every differential equation gives rise to WKB solutions computed by TR. In \cite{Deter,BBEnew}, it was proved that if the $\hbar$-dependent differential system satisfies some conditions, referred to as the Topological Type (TT) property, then the associated partition function has an expansion in $\hbar$ whose coefficients can be computed by TR.  A good understanding of the TT property is still missing but it could be proved that some interesting examples satisfy it. Among these examples, the simplest isomonodromic systems, i.e. the six Painlev\'e equations, were proved to be deformed by the introduction of a formal parameter $\hbar$ so that they satisfy the TT property \cite{P2,MarchalIwaki}. In the regime $\hbar \to 0$, the associated tau functions thus have a formal expansion given by TR. By the results of \cite{BDY,Eyn-intersection}, the coefficients of this formal expansion can thus be expressed in terms of intersection of tautological classes in the moduli space of Riemann surfaces generalizing Kontsevich-Witten theorem \cite{Kontsevich-KdV} obtained in the study of the Airy equation where the partition function is at the same time a KdV tau function and a generating function for Gromov-Witten invariants of the point.

For the understanding of the relation between integrable and isomonodromic systems, TR, knot invariants and CFTs, it is important to understand if this result can be generalized to other isomonodromic systems, namely, if one can deform a rational linear differential system in such a way that it satisfies the TT property. In this article, we prove that it is always possible to deform a rational linear system
$$
\frac{d }{dx}\Psi(x) = L(x) \Psi(x) \to \hbar \frac{d }{dx} \Psi(x,\hbar) = L(x,\hbar) \Psi(x,\hbar) 
$$
with $L(x) \in \mathfrak{sl}_2(\mathbb{C})(x)$, in such a way that 
\begin{itemize}
\item $L(x,\hbar)$ is equal to $L(x)$ for $\hbar = 1$.

\item The deformed equation $\hbar \frac{d}{dx} \Psi(x,\hbar)= L(x,\hbar) \Psi(x,\hbar)$ satisfies the TT property.
\end{itemize}

For doing so, as explained in section \ref{Section1}, we consider any generic $\mathfrak{sl}_2(\mathbb{C})$ valued rational function $L(x)$ as an element of a coadjoint orbit in a finite dimensional subspace of the $\mathfrak{sl}_2(\mathbb{C})$ loop algebra. In other words, after reduction, we consider it as a point in a symplectic space equipped with the corresponding Hitchin integrable system. We remind that this isospectral system has an autonomous Hamiltonian representation before de-autonomizing it to make it an isomonodromic system equipped with a non-autonomous Hamiltonian representation. This is done in section \ref{SectionHbarIntro} where we also give a prescription to introduce a formal parameter $\hbar$ giving rise to a family of isomonodromic systems deforming the original one. Finally, in section \ref{TRSection}, we state in Theorem \ref{main-theorem} our main result which is that the family of systems built in this way satisfies the TT property.
Hence, we prove that one can always deform a linear differential system characterized by $L(x) \in \mathfrak{sl}_2(\mathbb{C})(x)$ in such a way that the associated tau function has an expansion computed by TR. This generalizes the results known up to now on various examples in the context of isomonodromic systems. 

Moreover, our result allows to generalize the quantization procedure of \cite{Reconstruction}. Indeed, to any hyperelliptic curve of genus $0$, possibly with interior points in its Newton polytope, we can associate a linear system depending on $\hbar$ with the TT property. It means that one can build a quantum curve for any such classical hyperelliptic curve.

In section \ref{SectionExamples}, we present a few examples including the Painlev\'e equations as well as rank $2$ Fuchsian systems and arbitrary $\mathfrak{sl}_2(\mathbb{C})$ valued polynomials, the latter including some of the Painlev\'e hierarchies.

Since our approach uses mainly the Poisson structure of the loop algebra, we believe that it represents a first step towards a generalization to any isomonodromic system. It also allows to make the link with more geometrical approaches of the isomonodromic deformations as explained in section \ref{SectionConclusion}. In particular, we believe that the approach developed in this article should extend naturally to $L(x)\in \mathfrak{gl}_N(\mathbb{C})(x)$ with $N\geq 2$ and to $L(x)\in \mathfrak{g}(x)$ where $\mathfrak{g}$ is any complex simple Lie subalgebra of $\mathfrak{gl}_N(\mathbb{C})$.

\bigskip

{\bf Aknowledgements:} N.O. would like to thank Mattia Cafasso, John Harnad and Volodya Roubstov for helpful discutions on the subject as well as Universit\'e de Lyon for its hospitality.

O.M. would like to thank Universit\'e Lyon $1$, Universit\'e Jean Monnet and Institut Camille Jordan for material support. This work was supported by the LABEX MILYON (ANR-10-LABX-0070) of Universit\'e de Lyon, within the program « Investissements d'Avenir » (ANR-11-IDEX-0007) operated by the French National Research Agency (ANR).

\section{Topological recursion and WKB analysis}

\subsection{Compatible systems, correlators and loop equations}

In this article, we are interested in studying compatible systems of the form
$$
\left\{
\begin{array}{l}
\frac{\partial}{\partial x} \Psi(x,t) = L(x,t) \Psi(x,t) \cr
\frac{\partial}{\partial t} \Psi(x,t) = A(x,t) \Psi(x,t) \cr
\end{array}
\right. 
$$
where $L(x,t)$ and $A(x,t)$ belong to a simple Lie algebra $\mathfrak{g}\subset \mathcal{M}_N(\mathbb{C})$\footnote{The general setup maybe extended to an arbitrary semi-simple Lie algebra equipped with a faithful representation $\rho$ used to pushforward quantities in $\mathcal{M}_N(\mathbb{C})$ (See \cite{LoopLie2} for details)}. Moreover we assume that for all $t\in \mathbb{C}$,  the functions $x\mapsto L(x,t)$ and $x\mapsto A(x,t)$ are rational functions of $x \in \Sigma=\mathbb{P}^1$. Note that loop equations and correlators that will be defined below have been generalized in \cite{LoopLie2} for an arbitrary Riemann surface $\Sigma$. However for our purpose, we will restrict only to $\Sigma=\mathbb{P}^1$ in the present article. In particular, since $\mathbb{P}^1$ is a genus $0$ Riemann surface, its fundamental group and its corresponding action are trivial and this simplifies some definitions and properties presented in \cite{LoopLie2}. 

\medskip

To any solution $\Psi(x)$ of a linear differential equation
$$
\frac{\partial}{\partial x} \Psi(x) = L(x) \Psi(x)
$$
where $L(x)$ is a rational function of $x$ with values in a Lie algebra $\mathfrak{g}\subset\mathcal{M}_N(\mathbb{C})$, Eynard et al. \cite{LoopLie,LoopLie2,Deter,BBEnew} associate a set of elements
$$
M(x;\mathbf{E}):= \Psi(x) \mathbf{E} \Psi(x)^{-1}
$$
where $\mathbf{E} \in \mathfrak{h}$ is an arbitrary element of the Cartan subalgebra $\mathfrak{h}\subset \mathfrak{g}$.\footnote{Since the base field $\mathbb{C}$ is algebraically closed and the Lie algebra is finite dimensional then all Cartan subalgebras are conjugate under automorphisms of the Lie algebra, and thus are all isomorphic.} These elements satisfy the equation
$$
\frac{\partial}{ \partial x} M(x,\mathbf{E}) = \left[L(x),M(x,\mathbf{E})\right].
$$
Motivated by the correlation functions in random matrix models and CFT's, one can define a set of correlators associated to such a linear differential equation (Cf. \cite{LoopLie,LoopLie2,Deter,BBEnew}).
\bd [Correlation functions]\label{DefWn} For $n\geq1$, let the $n$-point correlators $W_n: \left(\mathbb{P}^1 \otimes \mathfrak{h} \right)^{n} \to \mathbb{P}^1$ be defined by
\small{$$
W_n(x_1 \otimes E_1 ,\dots, x_n \otimes E_n) := \left\{
\begin{array}{lr}
\Tr \left[L(x_1) M(x_1; E_1)\right] \, dx_1  &\,\text{ if }\, n=1 \cr
\frac{(-1)^n}{n} {\displaystyle \sum_{\sigma \in \mathcal{S}_n}} \frac{\Tr \left[  M(x_{\sigma(1)};E_{\sigma(1)}) M(x_{\sigma(2)};E_{\sigma(2)})\dots M(x_{\sigma(n)};E_{\sigma(n)})\right]
}{ {\displaystyle \prod_{i=1}^n} (x_{\sigma(i)}-x_{\sigma(i+1)})} dx_1\dots dx_n &\,\text{ if }\, n\geq 2 \cr
\end{array}
\right. 
$$}\normalsize{}
where $\mathcal{S}_n$ is the symmetric group of $n$ elements. 
\ed

In \cite{LoopLie2} (Theorem $4.3$) it was proved that they satisfy a set of loop equations, or $\mathcal{W}$-constraints. These loop equations greatly simplify in the case of $\mathfrak{g}=\mathfrak{sl}_N(\mathbb{C})$ presented in \cite{BBEnew} and even more in the case where $\mathfrak{g}=\mathfrak{sl}_2(\mathbb{C})$ developed earlier in \cite{Deter}. Since we will not make direct use of the loop equations in this article, we just refer the reader to Definition $2.6$ and Theorem $2.7$ of \cite{Deter} for loop equations arising in $\mathfrak{g}=\mathfrak{sl}_2(\mathbb{C})$.

\medskip

Let us just mention that generically, this set of loop equations does not have a unique solution. However, restricting the space of solutions by requiring additional properties for the correlators, known as the ``Topological Type property'', allows to prove uniqueness as well as solve explicitly this set of equations. Note that the correlators, the loop equations and the topological type property are defined for any rational linear differential system $\partial_x \Psi(x)=L(x)\Psi(x)$. In particular there is no need for a second compatible time differential equations to define them. On the other hand, as we show in this paper, the existence of such a compatible equation allows to naturally prove the topological type property which is usually hard to obtain otherwise.

\medskip

From the point of view of isomonodromic systems, i.e. when the differential equation in $x$ belongs to a compatible system
$$
\left\{
\begin{array}{l}
\frac{\partial}{\partial x} \Psi(x,t) = L(x,t) \Psi(x,t) \cr
\frac{\partial}{\partial t} \Psi(x,t) = A(x,t) \Psi(x,t) \cr
\end{array}
\right. ,
$$
one of the main motivations for the introduction of these correlators is the fact that they give access to the associated isomonodromic tau function. 

Generalizing the definition of Jimbo, Miwa and Ueno \cite{JMUI,JMUII}, Bertola and the first author \cite{BM}\footnote{A similar definition was given by \cite{CGL} for an arbitrary Riemann-Hilbert problem. We believe that these definitions are equivalent whenever the RH problem comes from an isomonodromic system.} defined a tau function $\tau_{BM}$ associated to such a system by identifying its variations with respect to the isomonodromic time $t$\footnote{In general, as we shall see in the following, one could have more than one isomonodromic time and the following formula should be valid for any of them. The existence of a solution to the corresponding set of equations was proven in \cite{BM}}.

Let us now recall how one can define such a tau function associated to an isomonodromic system. We shall consider the case $\mathfrak{g} = \mathfrak{sl}_N$ for $N \geq 2$ for this purpose.

Let $\Psi(x)$ be solution to an isomonodormic system and 
$$
\Psi(x) =  \Psi_-(x) x^S \Psi_+(x)
$$
be a Birkhoff factorization where $\Psi_-(x) \in x^{-1} SL_N[[x^{-1}]]$ (resp. $\Psi_+(x) \in SL_N[[x]]$) and $S \in GL_N$ is independent of $t$. One can then define the form \cite{BM}
$$
\omega_{BM} := \Res_{x \to \infty} \Tr \left(x^{-S} \Psi_-(x)^{-1} {d  \Psi_-(x) x^S \over dx} d\Psi_+(x) \Psi_+(x)^{-1}\right)
$$
where $d\Psi_+(x)$ denotes the differential of $\Psi_+(x)$ in the space of isomonodromic parameters.

It was proved in \cite{BM} that
\bt
The differential $\omega_{BM}$ is closed  and locally defines a tau function by
$$
d \ln \tau_{BM} = \omega_{BM}.
$$

\et

In \cite{BBEnew}, Berg\`ere, Borot and Eynard remarked that this tau function can be recovered from the knowledge of $W_1$ by integrating this definition with respect to the isomonodromic time $t$:
$$
\partial_t \ln \tau_{BM} = \oint_{x \in \mathcal{C} } W_1(x) f(x) dx
$$
where $\mathcal{C}$ (resp. $f(x)$) is a contour (resp. a function) to be specified depending on the system considered.

\subsection{Topological type property and topological recursion}

Let us now assume that the whole isomonodromic system depends on a formal parameter $\hbar$, taking the form
\beq\label{eq-isomono-system}
\left\{
\begin{array}{l}
\hbar \frac{\partial}{\partial x} \Psi(x,t,\hbar) = L(x,t,\hbar) \Psi(x,t,\hbar) \cr
\hbar \frac{\partial}{\partial t} \Psi(x,t,\hbar) = A(x,t,\hbar) \Psi(x,t,\hbar) \cr
\end{array}
\right. .
\eeq
Under some assumptions, the loop equations then have a unique solution in the space of formal series in $\hbar$ and this solution can be computed order by order by the topological recursion procedure of \cite{EO}. Furthermore, it gives a procedure to compute $\tau_{BM}$ order by order in $\hbar$ up to multiplicative factors independent of $t$ (the BM tau-function is defined up a global multiplicative constant). Using general properties of the topological recursion, the coefficients of this formal series expansion can be expressed in terms of intersection of cohomology classes in the Deligne-Mumford compactification of the moduli space of Riemann surfaces. We first recall the definition of a global classical spectral curve suitable for our purposes\footnote{The topological recursion has known many generalizations after \cite{EO}. In particular, the spectral curve may be replaced by some local data consisting of germs of differential forms \cite{DBOSS} but we only need the original simple version in this article.} and results about the topological recursion developed in \cite{EO}.

\bd[Classical spectral curve] A classical spectral curve  is composed of the following data.
\begin{enumerate}\item A compact  Riemann surface $\Sigma^{\text{cl}}$ equipped with a meromorphic function $x: \Sigma^{\text{cl}} \to \mathbb{P}^1$ such that the form $dx$ has simple zeroes at the branchpoints $\{b_i\}_{1\leq i\leq m}$ defined by $dx(b_i)=  0$. The degree of the map 
$$\begin{array}{ccccc}
x & : & \Sigma^{\text{cl}} & \to & \mathbb{P}^1 \\
 & & z & \mapsto & x(z) \\
\end{array}$$
is denoted by $N$. Thus, for any generic $x_0\in \mathbb{P}^1$, we may define the set of preimages $x^{-1}(x_0)=\{z^1(x_0),\dots,z^N(x_0)\}$ by choosing a labeling of the sheets of the cover.

\item A meromorphic form $\om_{0,1}$ on $\Sigma^{\text{cl}}$ holomorphic at the branch points $\{b_i\}_{1\leq i\leq m}$.

\item A symmetric bidifferential $\om_{0,2}$ on $\Sigma^{\text{cl}} \times \Sigma^{\text{cl}}$ with only double poles on the diagonal without residues.

\end{enumerate}
In the rest of the paper, we denote the set of branchpoints by $\mathcal{B} =\{b_i\}_{1\leq i\leq m}$. The number of branchpoints is denoted $m=|\mathcal{B}|\geq 1$.
\ed

We now recall that the topological recursion developed in \cite{EO} may be applied to any classical spectral curve under some genericity assumptions which we shall assume in the following.

\begin{definition}[Topological Recursion] \label{def-TT=property}For any classical spectral curve $(\Sigma^{\text{cl}},x,\om_{0,1},\om_{0,2})$, one may construct recursively symmetric $n$-forms $\left(\omega_{g,n}(z_1,\dots,z_n)\right)_{n\geq 1,g\geq 0}$ on $\left(\Sigma^{\text{cl}}\right)^n$ called Eynard-Orantin differentials and some symplectic invariants $\left(F^{(g)}\right)_{g\geq 0}$ using the topological recursion presented in \cite{EO}.
\end{definition}

Formulas defining the topological recursion can be found in \cite{EO}. Let us just remind that the recursion is done over the index $p=2g+n$ starting from $\omega_{0,1}$ and $\omega_{0,2}$ that are initial data coming from the definition of the classical spectral curve. The formulas involve residues at the branchpoints and it is known that the Eynard-Orantin differentials produced by the topological recursion are symmetric and may only have poles at the branchpoints (except $\omega_{0,2}$).

In order to apply the topological recursion procedure to the study of the asymptotics of a linear differential equation, the latter must satisfy the so-called Topological type properties. Since we only need to work with reductions of $\mathfrak{gl}_N(\mathbb{C})$, we consider the version of topological type properties for this algebra. 
Hence one can consider a basis of the corresponding Cartan sub-algebra given by diagonal matrices $(e_i)_{i=1}^N$ with only non-vanishing element $[e_i]_{j,k}:=\delta_{ij} \delta_{jk}$.

\bd[Topological Type Property]\label{DefTTproperty}
A system $\hbar \partial_x \Psi(x,\hbar)=L(x,\hbar)\Psi(x,\hbar)$, with a $N \times N$ matrix $L(x,\hbar)$, is said to satisfy the Topological Type property (TT property for short)  if the following conditions are met.
\begin{enumerate}

\item {\bf Genus 0 property:} The Riemann surface $\Sigma^{\text{cl}}$ given by the compactification of the $\hbar \to 0$ limit of the characteristic polynomial $\underset{\hbar\to 0}{ \lim} \det (yI_d - L(x,t)) = 0\} \subset \overline{\mathbb{C}}^{\,2}$ has genus $0$.

\item {\bf Formal expansion in $\hbar$:} The correlators admit a formal series expansion in $\hbar$.
\begin{eqnarray*}
W_1(x_1\otimes e_{a_1}) &=&  \sum_{k = -1}^\infty \hbar^k W_1^{(k)}(x_1\otimes e_{a_1})  := \sum_{k = -1}^\infty \hbar^k \td{W}_1^{(k)}(z^{a_1}(x_1))\cr
W_n(x_1\otimes e_{a_1},\dots,x_n\otimes e_{a_n})&=&  \sum_{k = 0}^\infty \hbar^k W_n^{(k)}(x_1\otimes e_{a_1},\dots,x_n\otimes e_{a_n})\cr
&:=& \sum_{k = 0}^\infty \hbar^k \td{W}_n^{(k)}(z^{a_1}(x_1) ,\dots,z^{a_n}(x_n)) \quad , \quad \forall\, n\geq 2,
\end{eqnarray*}
whose coefficients $\td{W}_n^{(k)}(z_1,\dots,z_n)dx(z_1)\dots dx(z_n)$ define symmetric $n$-forms on $\Sigma^{\text{cl}}$.


\item {\bf Parity property:} As functions of $\hbar$, the correlation functions have the parity property
$$
W_n(x_1\otimes E_{a_1},\dots,x_n\otimes E_{a_n}|-\hbar) = (-1)^n W_n(x_1\otimes E_{a_1},\dots,x_n\otimes E_{a_n}|\hbar) \,,\,\forall \, n\geq 1 .
$$

\item {\bf Pole property:} The $n$-forms $\td{W}_n^{(k)}(z_1,\dots,z_n)$ with $n\geq 1$ and $k\geq 0$ and $(n,k)\notin\{(1,0),(2,0)\}$ may only have singularities at the branchpoints of the classical spectral curve. Moreover these singularities may only be pole singularities. In other words, they are meromorphic functions on $\Sigma^{\text{cl}}$ with only pole singularities at the branchpoints. 

\item {\bf Leading order in $\hbar$:} For $n\geq 1$, the correlators satisfy $W_n = O(\hbar^{n-2})$ as functions of $\hbar$. In other words, for any $n\geq 2$, the coefficients $\left(W_n^{(k)}\right)_{0\leq k\leq n-1}$ are identically zero.  Note that combining properties $1$, $2$ and $5$ is equivalent to say that we have
$$ \td{W}_n(z_1,\dots,z_n)=\sum_{g=0}^{\infty} \td{W}_{g,n}(z_1,\dots,z_n)\hbar^{n-2-2g}.
$$

\item {\bf Identification of the two points function:} The $2$-form $\td{W}_{0,2}(z_1,z_2)dx(z_1)dx(z_2)$ defines an admissible symmetric form $\omega_{0,2}$  for the topological recursion.
\end{enumerate}
\ed

Then we have the following theorem \cite{LoopLie,BBEnew}.

\bt[Reconstruction with the Topological Recursion]
If the differential system $\hbar \partial_x \Psi(x)=L(x)\Psi(x)$ satisfies the TT property and $dx$ has only simple zeros on the classical spectral curve $\Sigma^{\text{cl}}$ then the coefficients of the correlation functions $\left(\td{W}_{g,n}(z_1,\dots,z_n)\right)_{n\geq 1,g\geq 0}$ identify with the corresponding Eynard-Orantin differentials $\left(\omega_{g,n}\right)_{n\geq 1,g\geq 0}$ computed from the application of the topological recursion of \cite{EO} to the classical spectral curve of the system with initial data given by $\omega_{1,0}(z)=\td{W}_{0,1}(z) dx(z)$ and $\omega_{2,0}(z_1,z_2)=\td{W}_{0,2}(z_1,z_2)dx(z_1)dx(z_2)$,
$$
\forall \,(z_1,\dots,z_n)\in\left(\Sigma^{\text{cl}}\right)^n\,:\, \td{W}_{g,n}(z_1 ,\dots, z_n )dx(z_1)\dots dx(z_n)=\omega_{g,n}(z_1,\dots,z_n).
$$
Moreover, if the differential system comes from an isomonodromic system of the form \eqref{eq-isomono-system} then the tau-function $\ln \tau_{\text{BM}}$ admits a formal series expansion in $\hbar$ of the form
\beqq \ln\tau_{\text{BM}}(t)=\sum_{g=0}^{+\infty} \tau^{(g)}(t)\hbar^{2g-2} \eeqq
where $\frac{d}{dt}\tau^{(g)}=\frac{d}{dt} F^{(g)}$ and $\left(F^{(g)}\right)_{g\geq 0}$ are the so-called symplectic invariant generated by the topological recursion applied to the classical spectral curve.  
\et

In the following sections, we shall explain how to build generic $\mathfrak{sl}_2(\mathbb{C})$-isomonodromic system and prove that they satisfy the TT property. This allows to compute a formal $\hbar$-expansion of the tau function and a formal $\hbar$ expansion of the correlators by application of the topological recursion. We believe that the construction presented below should extend to $\mathfrak{sl}_N(\mathbb{C})$ and to any simple Lie algebra over $\mathbb{C}$ (i.e $(A_n)_{n\geq 1}$, $(B_n)_{n\geq 1}$, $(C_n)_{n\geq 1}$, $(D_n)_{n\geq 2}$ and possibly the exceptional cases $E_6$, $E_7$, $E_8$, $F_4$ and $G_2$) but this is left for future works.

\br\label{rem-gl-vs-sl}
$\mathfrak{sl}_2$ is a rank one reduction of $\mathfrak{gl}_2$ obtained by vanishing of the trace. From the identification of the correlation functions with functions on the classical spectral curve, this means that one can work with the usual result for $\mathfrak{gl}_2$ by imposing the vanishing of the sum of the values of the functions considered over the two sheets of the classical spectral curve.

In particular, one only needs to prove the TT properties for the element $e_1$ of the $\mathfrak{gl}_2$ Cartan subalgebra. Indeed, values for $e_2$ are opposite to tgose for $e_1$ while values for the basis vector $e_1-e_2$ of the $\mathfrak{sl}_2$ Cartan subalgebra are equal to twice the corresponding values for $e_1$.
\er

\section{$SL_N$ isospectral deformation, integrable system and Poisson structure \label{Section1}}

The construction of isomonodromic systems admitting a WKB solution built by topological recursion relies heavily on the Poisson and symplectic structures on the space of flat connections on a Riemann surface as well as the associated Hitchin integrable system. Let us start by reviewing this structure from the $R$-matrix point of view \cite{Semenov-Tian-Shansky}.

\subsection{Isospectral systems and Poisson structure}

Let $\mathfrak{g}$ be a Lie algebra, and $\mathcal{C}$ be an oriented simple closed curve on $\mathbb{P}^1$. Let us denote by $U_+$ (resp. $U_-$) the region outside $\mathcal{C}$ (resp. inside $\mathcal{C}$).

One can associate to such a pair a loop space $\tilde{\mathfrak{g}}$ of smooth maps $L:\mathcal{C} \to \mathfrak{g}$ together with a polarization $\tilde{\mathfrak{g}} = \tilde{\mathfrak{g}}_+ \oplus \tilde{\mathfrak{g}}_-$ where $ \tilde{\mathfrak{g}}_- $ (resp. $\tilde{\mathfrak{g}}_+$) is the subalgebra of maps admitting an holomorphic extension to $U_-$ (resp. to $U_+$).

One can define an $\text{Ad}$-invariant inner product $<\cdot ,\cdot>:  \tilde{\mathfrak{g}} \times \tilde{\mathfrak{g}} \to \mathbb{C}$ by
$$
\forall\, (L_1,L_2) \in \tilde{\mathfrak{g}}\times  \tilde{\mathfrak{g}}\, , \; 
\left< L_1,L_2 \right> := \frac{1}{ 2 \pi i} \oint_{x \in \mathcal{C}} \Tr \left[L_1(x) \cdot L_2(x) \right] dx.
$$
This allows the identification of $\tilde{\mathfrak{g}}$ with its dual $\tilde{\mathfrak{g}}^*$  in such a way that the $\tilde{\mathfrak{g}}_{\pm}^*$ can be identified with $\tilde{\mathfrak{g}}_\mp$. 

The exponentiated group $\tilde{G}^*$ acts by coadjoint action through
$$
\forall \,(f,g) \in \tilde{\mathfrak{g}}^* \times \tilde{\mathfrak{g}} \, , \; \forall \,X \in \tilde{G}^* \, , \; 
\text{Ad}^*_X(f)(g) = \frac{1}{ 2 \pi i} \oint_{x \in \mathcal{C}} \Tr \left( [X,f] g\right).
$$

The polarization, which depends on the choice of a curve $\mathcal{C}$, also allows to define a second Lie algebra structure on the infinite dimensional space $\tilde{\mathfrak{g}}$ in addition to the one inherited from $\mathfrak{g}$. Using the classical $R$-matrix construction, one defines the bracket
$$
\forall \,(L_1,L_2) \in \tilde{\mathfrak{g}}\times \,\tilde{\mathfrak{g}} , \; \left[L_1,L_2\right]_R := \left[R(L_1),L_2\right]+\left[L_1,R(L_2)\right]
$$
where
$$
R:=\frac{1}{2} \left(P_+ - P_-\right)
$$
with $P_\pm$ being the projection operator to $U_\pm$. This defines a Lie-Poisson structure on $\tilde{\mathfrak{g}}^*$ through the bracket
$$
\forall \,(f,g) \in \tilde{\mathfrak{g}}^*\times \tilde{\mathfrak{g}}^*\, , \;  \{f,g\}_R(\mu) := \left<\mu,[df(\mu),dg(\mu)]_R\right>.
$$

%

Let us denote by $\mathcal{I}$ the set of spectral invariants, i.e. the set of $\text{Ad}^*$-invariant polynomials on $\tilde{\mathfrak{g}}^*$. 
The generalized Adler-Kostant-Symes (AKS) theorem in its $R$-matrix form
\cite{Semenov-Tian-Shansky} states that the elements of $\mathcal{I}$ Poisson commute. In addition, it provides the form of the Hamilton's equations. For $H \in \mathcal{I}$, the corresponding equation reads (see for example \cite{RMatrixHarnad2})
$$
\frac{d L}{dt} = [P_\sigma ( d H)  , L].
$$
where
\beq\label{Psigma}
P_\sigma := \frac{1}{2} \left[(1+\sigma)P_+ + (\sigma-1)P_- \right]
\eeq
for any $\sigma \in \mathbb{C}$ and $dH$ denotes the differential of $H$ evaluated at $L$ where the differential of a function $f$ is defined by the first order variation
$$ f(Y + \text{Ad}_X^* Y) = f(Y) + \text{Ad}_X^* Y(df) \text{ for any }X\in \tilde{G}^* .$$

\br
These Hamilton's equations are independent of $\sigma$. Indeed, since $P_+ + P_-=\text{Id}$: 
$$
P_\sigma = P_0 + \frac{\sigma}{2}  \text{Id} = R + \frac{\sigma}{2}  \text{Id} .
$$
Since $d H(L)$ commutes with $L$ for any element of $H \in \mathcal{I}$, one has
$$
\frac{d L}{dt} = [P_\sigma ( d H)  , L]  = [R( d H)  , L] + \frac{\sigma}{2} [ d H  , L] = [R( d H)  , L]
$$
for any $\sigma$.
\er

These equations give a set of isospectral deformations of a given operator $L \in \tilde{\mathfrak{g}}$, i.e. they imply that the spectrum of $L(x)$ is independent of $t$.

\br
It is important to notice that, even though Hamilton's equations are independent of the value of $\sigma$, the auxiliary matrix $A_H:=P_\sigma ( d H)$ is not! Hence, $\sigma$ can be used to fine tune the shape of this auxiliary matrix. All the procedure developed here depends explicitly on $\sigma$. However, for simplicity of notations, we shall not write the dependance of $A_H$ in $\sigma$ explicitly.
\er

From these equations, one can identify a set of Casimir functions and Hamiltonians of the system considered.

\subsection{$\mathfrak{sl}_N$  case, finite dimensional subspaces and generators of $\mathcal{I}$ \label{SectionHamiltonians}}
From now on, for simplicity, we shall consider $\mathcal{C}$ as a small contour encircling $x = \infty$. This allows to identify $\tilde{\mathfrak{g}}^+$ (resp. $\tilde{\mathfrak{g}}^-$) with elements of $\mathfrak{g}[[x]]$ (resp. $x^{-1} \mathfrak{g}[[x^{-1}]]$).
We also restrict to the case where $\mathfrak{g} = \mathfrak{sl}_N$ for $N \geq 2$. With this specialization, the ring of invariants $\mathcal{I}$ is generated by the coefficients of the expansion of $\Tr \left(L(x)^k\right)$ for $2\leq k \leq N$, namely
$$
\forall \,l \in \mathbb{Z} \, , \; \forall\, 2\leq k \leq N \, , \; h_{k,l} := \Res_{ x \to \infty} x^{-l-1} \, \Tr\left( L(x)^k\right) \, dx = \left<L(x)^k, x^{-l-1} \right>.
$$

We shall now consider finite dimensional subspaces of $\tilde{\mathfrak{g}}^*$. Namely, let us fix $n\geq 0$, a set of  points $\{a_1,\dots , a_n\}$  together with $a_0 := \infty$ as well as  integers $(r_\nu)_{1\leq \nu \leq n} \in \mathbb{N}^n$  and $r_0 \geq -1$.  We are finally given a fixed element $L_{0,r_0} \in \mathfrak{g}$.

Out of these data, denoting $r := \underset{\nu=0}{\overset{n}{\sum}} r_\nu$, let us define the subspace $\hat{\mathfrak{g}}^*$ of $\tilde{\mathfrak{g}}^*$ composed of elements with poles of order at most $r_\nu$ at $a_\nu$ by
\begin{itemize}
\item If $r_0\geq 0$, 
$$
\hat{\mathfrak{g}}^* := \left\{L(x) := \sum_{i=0}^{r_0} L_{0,i} x^i +  \sum_{\nu=1}^n \sum_{i=1}^{r_\nu} \frac{L_{\nu,i}}{  (x-a_\nu)^i} \, , \, (L_{\nu,k}) \in \mathfrak{g}^{r} \right\}
$$

\item If $r_0 = -1$,
$$
\hat{\mathfrak{g}}^* := \left\{L(x) :=   \sum_{\nu=1}^n \sum_{i=1}^{r_\nu} \frac{L_{\nu,i}}{ (x-a_\nu)^i} \, , \, (L_{\nu,k}) \in \mathfrak{g}^{r} | -\underset{\nu=1}{\overset{n}{\sum}} L_{\nu,1} =  L_{0,-1} \right\}
$$
where the residue at infinity, $-\underset{\nu=1}{\overset{n}{\sum}} L_{\nu,1} =  L_{0,-1}$, is fixed.
\end{itemize}

Let us now define a set of generators of $\hat{\mathcal{I}}$, the $Ad^*$-invariant functions on $\hat{\mathfrak{g}}^*$. They can be obtained by considering the coefficients of the partial fraction decomposition of each of the terms $\Tr \left( L(x)^k\right)$ to get\footnote{As we shall see, $H_{k,0,kr_0}$ might be vanishing if $L_{0,r_0}$ is not full rank.}
$$
\Tr \left( L(x)^k\right) = \sum_{i=0}^{k \, r_0} H_{k,0,i} x^i + \sum_{\nu=1}^n \sum_{i=1}^{k \, r_\nu} \frac{H_{k,\nu,i} }{ (x-a_\nu)^i}.
$$
 Thus, a set of generators of $\hat{\mathcal{I}}$ is given by the coefficients of these $Ad^*$-invariant functions defined by
 $$
\forall \,2\leq k \leq N \, , \; 1 \leq \nu \leq n \,  , \; 1 \leq i \leq k r_\nu \, , \; H_{k,\nu,i}  := - \frac{1}{ 2 \pi i} \oint_{ \infty} \frac{\underset{\mu \neq \nu}{\prod} (x-a_\mu)^{r_\mu}}{  \underset{\mu \neq \nu}{\prod} (a_\nu-a_\mu)^{r_\mu}}  (x-a_\nu)^{i-1} \Tr \left(L(x)^k\right) dx 
$$
and
$$
\forall\,  2\leq k \leq N \,   , \; 0 \leq i \leq k r_0 \, , \;  H_{k,0,i}:= \frac{1}{2 \pi i} \oint_{ \infty} x^{-i-1}\, \Tr \left(L(x)^k \right)dx .
$$

Indeed, $\frac{\prod_{\mu \neq \nu} (x-a_\mu)^{r_\mu} }{ \prod_{\mu \neq \nu} (a_\nu-a_\mu)^{r_\mu}}  (x-a_\nu)^{i-1} \Tr \left(L(x)^k\right) dx $ has poles only at $\infty$ and $a_\nu$. Thus one can move the integration contour and get
\bea \label{Ham} H_{k,\nu,i}&=&\frac{1}{2\pi i} \oint_{ a_\nu} \frac{\prod_{\mu \neq \nu} (x-a_\mu)^{r_\mu}  }{ \prod_{\mu \neq \nu} (a_\nu-a_\mu)^{r_\mu}}  (x-a_\nu)^{i-1} \Tr \left(L(x)^k\right) dx  \cr
&=&- \frac{1}{2\pi i} \oint_{ \infty} \frac{\prod_{\mu \neq \nu} (x-a_\mu)^{r_\mu}}{ \prod_{\mu \neq \nu} (a_\nu-a_\mu)^{r_\mu}}  (x-a_\nu)^{i-1} \Tr \left(L(x)^k\right) dx .\cr
&&
\eea

The associated Hamilton's equations read
$$
\frac{d L(x)}{ dt_{k,\nu,i}} = \left[A_{k,\nu,i},L(x)\right]
$$
where 
$$
A_{k,\nu,i} =\left\{
\begin{array}{l}
k \left[x^{-i-1}\, \left(L(x)\right)^{k-1}\right]_{+} = k \underset{x' \to \infty}{\Res} x'^{-i-1}\, \left(L(x')\right)^{k-1} \frac{dx' }{ (x'-x)} \quad \hbox{if} \quad \nu=0 \cr
- k \left[\frac{\prod_{\mu \neq \nu} (x-a_\mu)^{r_\mu} }{ \prod_{\mu \neq \nu} (a_\nu-a_\mu)^{r_\mu}}  (x-a_\nu)^{i-1} \, \left(L(x)\right)^{k-1}\right]_{+}
 \quad \hbox{otherwise}.\cr
\end{array}
\right.
$$
For $\nu \neq 0$, one can move the integration contour to get an expression localized around $a_\nu$,

\begin{eqnarray*}
A_{k,\nu,i}(x) &=&
- k \Res_{x' \to \infty} \frac{\prod_{\mu \neq \nu} (x'-a_\mu)^{r_\mu}  }{\prod_{\mu \neq \nu} (a_\nu-a_\mu)^{r_\mu}}  (x'-a_\nu)^{i-1} \, \left(L(x')\right)^{k-1} \frac{dx' }{ x'-x} \cr
&=& k  \frac{\prod_{\mu \neq \nu} (x-a_\mu)^{r_\mu} }{\prod_{\mu \neq \nu} (a_\nu-a_\mu)^{r_\mu}}  (x-a_\nu)^{i-1} \, \left(L(x)\right)^{k-1} \cr
&& \quad + k \Res_{x' \to a_\nu} \frac{\prod_{\mu \neq \nu} (x'-a_\mu)^{r_\mu} }{ \prod_{\mu \neq \nu} (a_\nu-a_\mu)^{r_\mu}}  (x'-a_\nu)^{i-1} \, \left(L(x')\right)^{k-1} \frac{dx' }{ x'-x} \cr
&=& k  \frac{\prod_{\mu \neq \nu} (x-a_\mu)^{r_\mu} }{ \prod_{\mu \neq \nu} (a_\nu-a_\mu)^{r_\mu}}  (x-a_\nu)^{i-1} \, \left(L(x)\right)^{k-1} + k  \left[ (x-a_\nu)^{i-1} \, \left(L(x)\right)^{k-1} \right]_{-,a_\nu}\cr
&&
\end{eqnarray*}
where the notation $[f(x)]_{-,a_\nu}$ stands for the polar part of $f(x)$ at $x = a_\nu$.

In addition, one can add to $A_{k,\nu,i}$ any matrix commuting with $L(x)$ without changing Hamilton's equations. Hence, we may remove the first contribution in the previous formula and get Hamilton's equations of the form
$$
\frac{d L(x)}{ dt_{k,\nu,i}} = \left[\tilde{A}_{k,\nu,i},L(x)\right]
$$
where 
$$
\tilde{A}_{k,\nu,i}  =  k \left[(x-a_\nu)^{i-1} \, \left(L(x)\right)^{k-1}\right]_{-,a_\nu}
$$
is the projection of $(x-a_\nu)^{i-1}\, \left(L(x)\right)^{k-1}$ to its polar part at $a_\nu$.

\br
For convenience, in some cases, it is useful to shift this auxiliary matrix by a matrix proportional to $L(x)$, i.e. taking $\sigma \neq 0$ in \eq{Psigma}. This is for example the case for recovering Painlev\'e IV equation.
\er

\subsection{Casimirs and symplectic leaves}
In order to describe the underlying integrable system, let us now restrict to symplectic leaves of $\hat{\mathfrak{g}}^*$ which are given by coadjoint orbits by first identifying a set of Casimir functions.

\bl
For any $\nu \neq 0$,
$H_{k,\nu,i}$ is a Casimir for $r_\nu (k - 1)+1\leq i\leq r_\nu k$ while $H_{k,0,i}$ is a Casimir for $r_0 (k - 1)\leq i\leq r_0 k$
\el

\proof{This follows from a direct computation. Indeed, for $\nu \neq 0$,  $\left(L(x)\right)^{k-1}$ being a polynomial of degree $r_\nu (k-1)$ in $(x-a_\nu)^{-1}$, $A_{k,\nu,i}$ is non-vanishing only for $1 \leq  i \leq r_\nu (k-1)$. In the same way, $A_{k,0,i}$ gives non-vanishing contributions for $0 \leq i \leq r_0 (k-1) - 1$.
}

\bigskip

By AKS theorem, the functions $H_{k,\nu,i}$ are in involution. Thanks to the preceding lemma, one can check that the number of non-Casimir Hamiltonians is equal to
$$
\sum_{k=2}^N \left[ r_0(k-1) + \sum_{\nu=1}^n r_\nu (k-1) \right] = r \frac{N (N-1)}{2}
$$
which matches with half the dimension of a generic symplectic leaf as expected for this integrable system\footnote{This is also true in the case $r_0 = -1$ since the residues $L_{\nu,1}$ are no longer independent.}.

\br
The point at infinity seems to play a specific role in this presentation. It indeed does since the contour $\mathcal{C}$ separates it from all other points. A different choice of contour would give rise to a different loop space and a different presentation of the same problem. It is expected that all the results should be independent of this choice of contour. This might be a manifestation of some invariance under the mapping class group of the surface considered.
\er

\subsection{Reduction}
The symplectic leafs described above are too large for our purpose. Because of the invariance of the correlators of the topological recursion under conjugation of $L(x)$ by a constant loop in $SL_N$, we would like to work with a symplectic manifold whose points represent coadjoint orbits under the $SL_N$ action. By fixing the value of $L_{0,r_0}$, we have already fixed part of the $SL_N$ action but not all of it\footnote{Remark that fixing $L_{0,r_0}$ could have been achieved by symplectic reduction by considering $L_{0,r_0}$ as a moment map.}.

\sloppy{Indeed, according to section \ref{SectionHamiltonians}, the $N-1$ flows corresponding to Hamiltonians $\left(H_{k,0,r_0(k-1)-1}\right)_{k\in\llbracket 2,N\rrbracket}$ lead to  auxiliary matrices $\left(A_{k,0,r_0(k-1)-1}\right)_{k\in\llbracket 2,N\rrbracket}$ independent of $x$ and commuting with $L_{0,r_0}$. The corresponding Hamiltonian flows generate conjugations by elements of the stabilizer $\text{Stab}_{L_{0,r_0}}$ of $L_{0,r_0}$\footnote{Once again, extra care has to be taken for the cases $r_0 \in \{-1,0\}$ but the same conclusion holds in these cases.}.}

Working modulo the coadjoint action of $SL_N$, one may fix the values of the corresponding Hamiltonians by symplectic reduction.

This reduction leads to a reduced symplectic space $\hat{\mathfrak{g}}^{\text{red}} := \hat{\mathfrak{g}}^* \sslash \text{Stab}_{L_{0,r_0}}$ of dimension
$$
D^{\text{red}} = D - 2 (N-1) = (r N-2) (N-1) \overset{\text{def}}{=} 2 g
$$
where $g$ is equal to the genus of the spectral curve $\text{det}(y-L(x)) =0 \subset \mathbb{C}^2$, i.e. to the dimension of its Jacobian, giving rise to the usual Hitchin Hamiltonian system. This reduced space is obtained by fixing a level set for the Hamiltonians $\left(H_{k,0,r_0(k-1)-1}\right)_{k\in\llbracket 2,N\rrbracket}$ and modding out by the action of the stabilizer $\text{Stab}_{L_{0,r_0}}$ of $L_{0,r_0}$.
 
The symplectic reduction implies that the remaining auxiliary matrices should have a contribution coming from the projection to these orbits leading to a shift
$$
A_{k,l} \mapsto A_{k,l} +\sum_{i=2}^N \beta_i A_{i,0,r_0(i-1)-1}
$$
for some constants $(\beta_2,\dots,\beta_N) \in \mathbb{C}^{N-1}$.

\subsection{Spectral Darboux coordinates in the $\mathfrak{sl}_2$ case \label{DarbouxSection}}
When restricting to such an orbit, one is left with a Hitchin integrable system of dimension $2g$ which is a fibration above a base parametrized by the spectral invariants $H_{k,\nu,l}$ with fiber equal to the Jacobian of the spectral curve considered (i.e. a point in the fiber is a line bundle over the spectral curve, giving the Abelian property of the fibers).

In the present article, restricting to the $\mathfrak{sl}_2(\mathbb{C})$ case, we shall not need this geometric picture since one can make all the computations explicit. However, the knowledge of this integrable system still provides a guiding principle for our work and is very useful for the generalization of this procedure to higher rank algebras as well as base curves of different topologies.

It is thus useful to remind how one can equip a generic symplectic leaf with spectral Darboux coordinates \cite{AHH2,Darboux} allowing to identify our examples with well-known Lax pairs found in the literature. In the following, we shall often use a representation of $\mathfrak{sl}_2$ by $2 \times 2$ matrices with a basis given by
$$
\sigma_3 := \begin{pmatrix}
1 & 0 \cr
0 & - 1\cr
\end{pmatrix}=e_1-e_2
\quad , \quad 
\sigma_+ :=  \begin{pmatrix}
0 & 1 \cr
0 & 0\cr
\end{pmatrix}
 \quad \hbox{and} \quad 
\sigma_- :=  \begin{pmatrix}
0 & 0 \cr
-1 & 0\cr
\end{pmatrix} . 
$$

These Darboux coordinates are easy to describe in the $\mathfrak{sl}_2(\mathbb{C})$ case and we shall restrict to this case in the present section. Considering  $L_{0,r_0}$ such that the component $\left(L_{0,r_0}\right)_{2,1}$ is vanishing, $\left[L(x)\right]_{2,1}$ has $g$ zeroes located at $x = q_i$, $i\in\llbracket1,g\rrbracket$. At these values, $L(x)$ takes the form
$$
L(q_i) =  \begin{pmatrix}
p_i & \left[L(q_i)\right]_{1,2} \cr
0 & -p_i \cr
\end{pmatrix}
$$
where $p_i = \left[L(q_i)\right]_{1,1}$ is an eigenvalue of $L(q_i)$.  The set $(p_i,q_i)_{i=1}^g$ form a system of Darboux coordinates on the symplectic leaf going through $L(x)$. 

Since $p_i$ is an eigenvalue of $L(q_i)$, $(q_i,p_i)$ satisfy $\det ( p_i-L(q_i)) = 0$  and such a pair defines a point on the spectral curve. Fixing a point in the Hitchin base, i.e. fixing the spectral curve, the set $(p_i,q_i)_{i=1}^g$ defines $g$ points on the spectral curve, i.e. a point in its Jacobian which is the fiber of the integrable system above the corresponding point in the base.

Note that the element $L(x)$ can  be recovered by Lagrange interpolation.

Since  $(p_i,q_i)_{i=1}^g$  provide Darboux coordinates, they satisfy the evolution equations
\beq\label{evolution1}
\forall \,i\in\llbracket 1,g \rrbracket\, , \; \left\{
\begin{array}{l}
\frac{\partial q_i }{ \partial t_{k,\nu,l}} = \frac{\partial H_{k,\nu,l}({\bf{p}}, {\bf{q}} ) }{ \partial p_i} \cr
\frac{\partial p_i }{ \partial t_{k,\nu,l}} = - \frac{\partial H_{k,\nu,l}({\bf{p}}, {\bf{q}} ) }{\partial q_i} \cr
\end{array}
\right.
\eeq
for any triple $(k,\nu,l)$ such that $H_{k,\nu,l}$ is not a Casimir.

\subsection{Representative of a reduced orbit in the $\mathfrak{sl}_2$ case\label{SL2Action}}
In the following, it will sometimes be convenient to represent a reduced orbit by one of its representatives. Since all the results presented in this article are invariant under the action of  $\text{Stab}_{L_{0,r_0}}$, we may choose any representative we need. For the derivation of the topological type property, one shall make the following choice. 
\begin{enumerate}
\item In the case where $L_{0,r_0} = \sigma_3$, we fix the value of an additional element $L_{\nu,k}$ to be of the form
$$
L_{\nu,k}=
\begin{pmatrix}
U & 1 \cr
V & -U \cr
\end{pmatrix} .
$$

\item In the case where $(L_{0,r_0}) = \sigma_+$, we fix the value of an additional element $L_{\nu,k}$ to be of the form
$$
L_{\nu,k}=
\begin{pmatrix}
0 & V \cr
1 & 0 \cr
\end{pmatrix} .
$$
\end{enumerate}

\begin{remark}The previous choices of representatives are possible only if the second term $L_{\nu,k}$ has a non-zero off-diagonal value in the $\sigma_3$ case (in which case after a potential transposition leaving $\sigma_3$ invariant, we may always set it to the upper-right entry) or a non-zero lower-left entry in the $\sigma_+$ case. However, in these degenerate cases, the degree of the Poisson manifold decreases and thus the generic theory presented in this paper fails to apply directly. Thus, in the rest of the paper, we will only consider cases where the orbits admit a representative of the form presented above.
\end{remark}

\begin{remark} In practice, when $r_0\geq 1$, one often chooses to fix the shape of $L_{0,r_0-1}$ in this way. In this article, this choice will be particularly useful in the study of the potential pole singularity at $x=\infty$ presented in appendix \ref{AppendixA}.
\end{remark}

\section{Building an $\hbar$-deformed isomonodromic system from an isospectral one\label{SectionHbarIntro}}

In the preceding section, we reminded how to build a symplectic leaf going through a given point $L(x)$ whose dimension equals the genus of the associated spectral curve such that the corresponding Hamiltonian flows are isospectral. In this section, we show that one can associate to any such isospectral system a (non-unique) isomonodromic system by replacing the autonomous Hamiltonians by non-autonomous ones. We also prove that one can introduce a formal parameter $\hbar$ rescaling Hamilton's equations around the point $\hbar = 0$ in the base of the integrable system where the fiber is completely degenerate.

\subsection{Isomonodromic systems from non-autonomous Hamiltonians}
Let us remind how to build an isomonodromic system by de-autonomization of an isospectral one. If a time $t$ preserves the spectrum of a matrix $L$, the compatibility of the system
$$
\left\{
\begin{array}{l}
L(t) \psi(t) = \mu \psi \cr
\frac{\partial}{\partial t} \psi(t) = A(t) \psi(t) \cr
\end{array}
\right.
$$
takes the Lax form
\beq \label{Lax-isospectral}
\frac{\partial L}{\partial t} = \left[A(t), L\right]
\eeq
which is the form of the Hamiltonian flows studied up to now.

We are actually interested in the compatibility of a system of the form
$$
\left\{
\begin{array}{l}
\frac{\partial}{ \partial x} \Psi(x,t)  = L(x,t) \Psi(x,t)  \cr
\frac{\partial }{ \partial t} \Psi(x,t) = A(x,t) \Psi(x,t) \cr
\end{array}
\right.
$$
which reads
\beq \label{Lax-isomonodromic}
\frac{\partial L}{ \partial t} - \frac{\partial A}{ \partial x} = \left[A, L\right].
\eeq

One can trade the former compatibility condition \eq{Lax-isospectral} to the latter \eq{Lax-isomonodromic} by replacing an autonomous system by a non-autonomous one. Namely, let us assume that one makes the matrix $L(x,t,a)$ depend on a parameter $a \in \mathcal{I}$ such that 
$$
\left. \frac{\partial L}{ \partial a}\right|_{a=t} = \frac{\partial A}{ \partial x}.
$$
Then the former compatibility condition includes both the Hamiltonian flow with respect to $t$ and the variation in $t$ coming from the $a=t$ condition and gives the isomonodromic compatibility condition thanks to the relation between $A$ and $L$. Let us now prove that one can build such an isomonodromic system in the $N=2$ case.

\subsection{Isomonodromic systems with linear auxiliary matrices \label{SectionLinearAuxMat}}
To derive an isomonodromic system admitting the topological type property for $N=2$, it is enough to have one isomonodromic time leading to an auxiliary matrix of the form 
$$
A(x) = \frac{M x + B }{p(x)}
$$
where $p(x) \in \mathbb{C}[x]$ is a $\mathbb{C}$-valued polynomial in $x$ while $(M,B) \in \mathfrak{sl}_2(\mathbb{C})\times \mathfrak{sl}_2(\mathbb{C})$ do not depend on $x$. In this section, we prove that it is always possible to derive such an isomonodromic deformation with $\sigma=0$ in \eq{Psigma}. We will see in some of the examples of section \ref{SectionExamples} that it is sometimes more convenient to consider other values of $\sigma$ but for the purpose of the article we only need to prove that there exists at least one suitable isomonodromic deformation.

For this purpose, one can consider different possible deformations of  an element $L \in \hat{\mathfrak{g}}^{\text{red}}$ depending on its pole structure. 

\begin{enumerate}
\item If there exists $\nu \neq 0$ such that $r_\nu=1$, one can see that
$$
A_{2,\nu,1}(x) = 2 \frac{ L_{\nu,1}}{ (x-a_\nu)}.
$$
Hence,
$$
\frac{1}{2} \frac{\partial A_{2,\nu,1} (x) }{ \partial x }= - \frac{ L_{\nu,1}}{ (x-a_\nu)^2} = \frac{\partial L(x) }{ \partial a_\nu}.
$$

One thus gets an isomonodromic system with identifying $t = \frac{1}{2}t_{2,\nu,1} = a_\nu$ leading to
$$
\left\{
\begin{array}{l}
\frac{\partial}{ \partial x} \Psi(x,t)  = L(x,a_\nu = t) \Psi(x,t)  \cr
\frac{\partial }{ \partial t} \Psi(x,t) = \left( \frac{ L_{\nu,1}}{  (x-t)} + \underset{k=2}{\overset{N}{\sum}}\beta_k A_{k,0,r_0(k-1)-1} \right) \Psi(x,t) \cr
\end{array}
\right. 
$$
where one has taken into account the reduction for defining the auxiliary matrix.

\bigskip

\item If there exists $\nu \neq 0$ such that $r_\nu\geq2$, one can define
$$
\frac{A_{2,\nu,r_\nu}}{ 2} = \frac{L_{r_\nu} }{(x-a_\nu)}
$$
in such a way that
$$
\frac{1}{2}\frac{\partial A_{2,\nu,r_\nu}}{\partial x} = - \frac{L_{r_\nu}}{ (x-a_\nu)^2}.
$$
This allows to define an isomonodromic system through the identification $t = \frac{1}{2}t_{2,\nu,r_\nu}$ 
$$
\left\{
\begin{array}{l}
\frac{\partial }{ \partial x} \Psi(x,t)  = \left[L(x) - t \frac{L_{r_\nu} }{ (x-a_\nu)^2} \right]  \Psi(x,t)  \cr
\frac{\partial }{ \partial t} \Psi(x,t) = \left(\frac{L_{r_\nu}}{ (x-a_\nu)} +  \underset{k=2}{\overset{N}{\sum}} \beta_k A_{k,0,r_0(k-1)-1} \right) \Psi(x,t) \cr
\end{array}
\right. .
$$

\br
The condition $r_\nu \geq 2$ is necessary in order to introduce the deformation $t \frac{L_{r_\nu}}{(x-a_\nu)^2}$ without changing the value of $r_\nu$. The preceding deformation allows to consider the simple pole case $r_\nu = 1$.
\er

\bigskip

\item Finally, if $r_0 \geq 2$ for $\nu =0$, one has
$$
\frac{A_{2,0,r_0-2}(x) }{ 2} = L_{0,r_0-1} + L_{0,r_0} x 
$$
leading to 
$$
\frac{1 }{2} \frac{\partial A_{2,0,r_0-2}(x) }{ \partial x}  = L_{0,r_0}.
$$
One can thus define an isomonodromic system by identifying $t = \frac{1}{2}t_{2,0,r_0-2}$
\beq \label{ACase3}
\left\{
\begin{array}{l}
\frac{\partial }{ \partial x} \Psi(x,t)  = \left[L(x) + t L_{0,r_0} \right]  \Psi(x,t)  \cr
\frac{\partial }{ \partial t} \Psi(x,t) = \left(  L_{0,r_0-1} + L_{0,r_0} x + \underset{k=1}{\overset{N}{\sum}} \beta_k A_{k,0,r_0(k-1)-1} \right) \Psi(x,t) \cr
\end{array}
\right.
\eeq
\end{enumerate}

With these three possibilities, one exhausts all possible cases for $N=2$ apart from the cases $n=0$ and $r_0\in\{0,1\}$.  These cases are irrelevant from the integrable systems perspective developed here since the corresponding dimension of the reduced phase space $D^{\text{red}} = (r_0 N-2) (N-1)$ is non positive. Moreover, the case $(n,r_0)=(0,0)$ is equivalent to $L(x)=L_{0,0}$ independent of $x$ for which there exist explicit solutions $\Psi(x)$ and no need for this theory. Cases for $(n,r_0)=(0,1)$, i.e. $L(x)=L_{0,0}+L_{0,1}x$ correspond to either the Airy (if $\text{rank}(L_{0,1})=1$) or the Hermite-Weber (if $\text{rank}(L_{0,1})=2$) systems that are well-known to satisfy the TT property (as a consequence of \cite{Reconstruction} since the corresponding Newton polytopes of the classical spectral curves have no interior points, or as specific limits of the Painlev\'e I or II Lax systems as explained in \cite{P2}). Note that for $N\geq 3$, cases with $n=0$ and $r_0=1$ would provide a positive dimension of the reduced phase space and thus would require a similar analysis as the one developed above.

\bt
The integrable system defined on the coadjoint orbit through any $\mathfrak{sl}_2$ valued rational function $L(x)$ can be deformed into an isomonodromic system
$$
\left\{
\begin{array}{l}
\frac{\partial}{ \partial x} \Psi(x,t)  =L(x,t) \Psi(x,t)  \cr
\frac{\partial }{ \partial t} \Psi(x,t) = A(x,t) \Psi(x,t) \cr
\end{array}
\right. 
$$
where $A(x) = \frac{M x + B }{p(x)}$ with $p\in \mathbb{C}[X]$ and $(M,B)\in \left(\mathfrak{sl}_2(\mathbb{C})\right)^2$ and $L(x,t=0) = L(x)$.
\et

\br
It is worth noticing that the different cases considered are non-exclusive. If an element $L(x)$ falls into more than one of the cases presented above, all the corresponding matrices $A$ are automatically compatible and one may choose to pick any of them for its purposes. In our case, we will only need to take a single matrix to prove the topological type property, and all other compatible matrices will not be considered in our procedure.
\er

\subsection{Introduction of $\hbar$.}\label{section-introduction-h}
One shall now introduce a formal parameter $\hbar$ in order to replace Hamilton's equation  for the autonomous system by
$$
\hbar \frac{\partial L(x,\hbar) }{\partial t} = \left[A(x,\hbar),L(x,\hbar)\right]
$$ 
and thus the compatibility condition for the non-autonomous one by
$$
\hbar \frac{\partial L(x,t,\hbar) }{ \partial t} - \hbar \frac{\partial A(x,t,\hbar)}{ \partial x} = \left[A(x,t,\hbar),L(x,t,\hbar)\right].
$$ 

One can easily obtain such a deformation by finding a rescaling of the variable $x$, the Hamiltonians $H_{k,\nu,i}$ and the times $t_{k,\nu,i}$ by some $\hbar^{d_q}$ factor, with $d_q \in \mathbb{Q}$, such that $t \cdot H = O(\hbar^2)$, $L_{0,r_0}$ is kept independent of $\hbar$ and each component of $L(x)$ as well as $\Tr \left[L(x)\right]^2$ are homogenous in $\hbar$.

For this purpose, one rescales $x \to  \hbar^{d_x} \, x$, $H_{k,\nu,i} \to \hbar^{ d_{k,\nu,i}} H_{k,\nu,i}$ and $t_{k,\nu,i} \to \hbar^{ d_{t_{k,\nu,i}}} t_{k,\nu,i}$. 

\bt
For any $L \in \tilde{\mathfrak{sl}_2}$ giving rise to an isomonodromic system as in section \ref{SectionLinearAuxMat}, there exists a set of exponents $(d_x,d_{k,\nu,i},d_{t_{k,\nu,i}})$ such that the compatibility equations are transformed into 
$$
\hbar \frac{\partial L(x,t,\hbar)}{ \partial t} - \hbar \frac{\partial A(x,t,\hbar) }{ \partial x} = \left[A(x,t,\hbar),L(x,t,\hbar)\right]
$$ 
by a rescaling of $t\to \hbar^{d_t}t$ and 
\footnotesize{$$
(x,a_\nu,H_{2,0,i},H_{2,\nu,i}) \to 
\left\{
\begin{array}{l}
\left(\hbar^{ \frac{1}{ r_0+1}} x, \hbar^{ \frac{1}{ r_0+1}} a_\nu,\hbar^{ \frac{2r_0-i}{ r_0+1}}H_{2,0,i}, \hbar^{ \frac{2r_0+i}{ r_0+1}}H_{2,\nu,i}\right)\,\text{ if }\, \text{rank}(L_{0,r_0})=2 \cr
\left(\hbar^{ \frac{2} { 2r_0+1}} x, \hbar^{\frac{2} { 2r_0+1}} a_\nu,\hbar^{ \frac{2(2 r_0-i-1)}{ 2r_0+1}}H_{2,0,i}, \hbar^{ \frac{2(2 r_0+i-1)}{2r_0+1}}H_{2,\nu,i}\right)\,\text{ if }\, \text{rank}( L_{0,r_0})=1\cr
\end{array}
\right. .
$$}\normalsize{}
\et

\proof{
The proof follows from the explicit construction of such deformations in the different cases introduced for building the isomonodromic system from the isospectral one. These are not the only possible ways of introducing $\hbar$ but it shows that it is always possible to find at least one. One shall only give the explicit computation in the case $r_0 \geq 0$. The case $r_0 = -1$ can be treated in a similar way.

{\bf{Case 1.}}

Let us first consider the case $\nu\neq 0$ and $r_\nu=1$. In this case, according to section \ref{SectionLinearAuxMat}, one considers $t = a_\nu$ as isomonodromic time. It has the same degree as $x$. The associated Hamiltonian is $H_{2,\nu,1}$.
We first compute $d_{2,\nu,1}$. Its value depends on the leading order $L_{0,r_0}$. Indeed, the degree in $x$ of $\Tr \left( L(x)^2\right)$ depends on  the rank of $L_{0,r_0}$. For this purpose, one shall consider the cases where $L_{0,r_0}$ as rank 1 or 2 separately. The degree of $\Tr \left(L(x)^2\right)$ is equal to $d_x \, (2r_0-1)$ in the first case and $2r_0 \, d_x$ in the second case.

The homogeneity constraint for $\Tr \left(L(x)^2\right)$ imposes that
$$
d_{2,\nu,1} = \left\{
\begin{array}{l}
{d_x} \left(2r_0 + 1\right)  \,\text{ if }\, \text{rank}( L_{0,r_0})=2 \cr
{d_x} \left(2r_0 \right) \,\text{ if }\, \text{rank}( L_{0,r_0})=1 \cr
\end{array}
\right. .
$$
The fact that the degrees of $t$ and $H$ sum up to 2 is thus equivalent to
$
2 = d_x+ d_{2,\nu,1}$ which leads to 
$$
{d_x} = 
\left\{
\begin{array}{l}
\frac{1}{  r_0 + 1}   \,\text{ if }\, \text{rank}(L_{0,r_0})=2 \cr
\frac{2 }{ 2r_0 +1} \,\text{ if }\, \text{rank}(L_{0,r_0})=1 \cr
\end{array}
\right. .
$$
One finally gets the associated deformed system through the rescaling
\footnotesize{$$
\left(t,x,a_\nu,H_{2,0,i},H_{2,\nu,i}\right) \to 
\left\{
\begin{array}{l}
\left(\hbar^{ \frac{1}{r_0+1}} t,\hbar^{\frac{1}{ r_0+1}} x, \hbar^{ \frac{1}{ r_0+1}} a_\nu,\hbar^{ \frac{2 r_0-i}{r_0+1}}H_{2,0,i}, \hbar^{ \frac{2 r_0+i}{r_0+1}}H_{2,\nu,i}\right)  \,\text{ if }\, \text{rank}(L_{0,r_0})=2 \cr
\left(\hbar^{\frac{ 2}{2r_0+1}} t,\hbar^{ \frac{2}{2r_0+1}} x, \hbar^{ \frac{2}{2r_0+1}} a_\nu,\hbar^{ \frac{2(2 r_0-i-1)}{ 2r_0+1}}H_{2,0,i}, \hbar^{ \frac{2(2 r_0+i-1) }{ 2r_0+1}}H_{2,\nu,i}\right) \,\text{ if }\, \text{rank}(L_{0,r_0})=1\cr
\end{array}
\right.
$$}\normalsize{}
Note that all exponents are rational numbers.

\bigskip

{\bf{Case 2:
}}
Let us now consider the case when $r_ \nu \geq 2$ for some $\nu \neq 0$.
The general procedure defines $t = t_{2,\nu,r_\nu}$ as isomonodromic time. From the homogeneity of $L(x)$ and with similar computations as above, one finds that
$$
d_{t_{2,\nu,r_\nu}} = (2-r_\nu) {d_x}.
$$
Then the homogeneity of $\Tr \left(L(x)^2\right)$ gives
$$
d_{2,\nu,i} = \left\{
\begin{array}{l}
{d_x} \left(2r_0 + i\right) \,\text{ if }\, \text{rank}(L_{0,r_0})=2 \cr
{d_x}\left(2r_0 -1+ i \right) \,\text{ if }\, \text{rank}(L_{0,r_0})=1 \cr
\end{array}
\right. .
$$
$d_{t_{2,\nu,r_\nu}}+ d_{2,\nu,r_\nu}  = 2$ then  implies
$$
{d_x} = \left\{
\begin{array}{l}
\frac{1 }{r_0 + 1} \,\text{ if }\, \text{rank}(L_{0,r_0})=2 \cr
\frac{2 }{2r_0 +1}  \,\text{ if }\, \text{rank}(L_{0,r_0})=1 \cr
\end{array}
\right. .
$$
One gets the associated systems by rescaling 
\footnotesize{$$
\left(t,x,a_\nu,H_{2,0,i},H_{2,\nu,i}\right) \to 
\left\{
\begin{array}{l}
\left(\hbar^{ \frac{2-r_\nu}{ r_0+1}} t,\hbar^{ \frac{1}{ r_0+1}} x, \hbar^{\frac{1}{ r_0+1}} a_\nu,\hbar^{ \frac{2 r_0-i}{ r_0+1}}H_{2,0,i}, \hbar^{\frac{2 r_0+i}{r_0+1}}H_{2,\nu,i}\right)  \,\text{ if }\, \text{rank}(L_{0,r_0})=2 \cr
\left(\hbar^{ \frac{2 (2-r_\nu) }{2r_0+1}} t,\hbar^{ \frac{2}{ 2r_0+1}} x, \hbar^{ \frac{2}{2r_0+1}} a_\nu,\hbar^{ \frac{2(2 r_0-i-1) }{ 2r_0+1}}H_{2,0,i}, \hbar^{ \frac{2(2 r_0+i-1) }{ 2r_0+1}}H_{2,\nu,i}\right)  \,\text{ if }\, \text{rank}( L_{0,r_0})=1 \cr
\end{array}
\right.
$$}\normalsize{}
Note again that all exponents are rational numbers.

\bigskip

{\bf{Case 3.}}
Let us finally consider the case
$$
L(x) = \sum_{i=0}^{r_0} L_{0,i} x^i + \sum_{\nu=1}^n \sum_{i=1}^{r_\nu} \frac{L_{\nu,i}}{ (x-a_\nu)^i}
$$
with $r_0 \geq 2$.

The isomonodromic time $t = \frac{1}{2}t_{2,0,r_0-2}$.  The homogeneity condition for $\Tr \left(L(x)^2\right)$ implies that 
$$
d_{2,0,i} = \left\{
\begin{array}{l}
{d_x} \left(2r_0 - i\right)  \,\text{ if }\, \text{rank}(L_{0,r_0})=2 \cr
{d_x} \left(2r_0 -1- i \right) \,\text{ if }\, \text{rank}(L_{0,r_0})=1 \cr
\end{array}
\right. .
$$
On the other hand, from the homogeneity of $L(x)$, one sees that $t$ is of order $x^{r_0}$ leading to
$$
d_{t_{2,0,r_0-2}} = r_0 \, {d_x}.
$$
The requirement $t . H=O\left(\hbar^2\right)$ thus reads
$$
2 = d_{t_{2,0,r_0-2}} + d_{2,0,r_0-2} = \left\{
\begin{array}{l}
{d_x} (2r_0+2)  \,\text{ if }\, \text{rank}(L_{0,r_0})=2 \cr
{d_x} (2r_0+1)  \,\text{ if }\, \text{rank}(L_{0,r_0})=1 \cr
\end{array}
\right. ,
$$
i.e.
$$
{d_x} = \left\{
\begin{array}{l}
\frac{1}{r_0+1}  \,\text{ if }\, \text{rank}(L_{0,r_0})=2 \cr
\frac{2}{2r_0+1}  \,\text{ if }\, \text{rank}(L_{0,r_0})=1 \cr
\end{array}
\right. .
$$
One gets the associated systems by rescaling 
\footnotesize{$$
(t,x,a_\nu,H_{2,0,i},H_{2,\nu,i}) \to 
\left\{
\begin{array}{l}
\left(\hbar^{ \frac{r_0 }{r_0+1}} t,\hbar^{\frac{1}{ r_0+1}} x, \hbar^{\frac{1}{r_0+1}} a_\nu,\hbar^{\frac{2r_0-i }{ r_0+1} }H_{2,0,i}, \hbar^{\frac{2r_0+i }{ r_0+1} } H_{2,\nu,i}\right) \,\text{ if }\, \text{rank}(L_{0,r_0})=2 \cr
\left(\hbar^{ \frac{2 r_0 }{ 2r_0+1}} t,\hbar^{\frac{2 }{ 2 r_0+1}} x, \hbar^{\frac{2}{2 r_0+1}} a_\nu,\hbar^{\frac{4r_0-2 i-2}{ 2r_0+1} }H_{2,0,i}, \hbar^{\frac{4r_0+2 i -2}{ 2 r_0+1} } H_{2,\nu,i}\right)  \,\text{ if }\, \text{rank}(L_{0,r_0})=1 \cr
\end{array}
\right. .
$$}\normalsize{}
}

\br 
If we have proved that one can always $\hbar$-deform such an isomonodromic system, there exist many other ways to do it. Examples will be given by some of the Painlev\'e equations. In any case, the procedure for determining the introduction of $\hbar$ always follows from some homogeneity conditions as well as the constraint for the order of the product $t H$.

In the introduction of $\hbar$, one can consider  $x$ as a dynamical variable, i.e. changing the value of $\hbar$ changes the base curve, or keep it independent of $\hbar$. We always considered the first possibility here but we shall see in examples that another choice can be made.
\er

After performing this non-autonomous deformation and introducing $\hbar$, the dynamics of the system is described by
\beq\label{evolution2}
\forall \,i\in\llbracket 1,g\rrbracket \, , \; \left\{
\begin{array}{l}
\hbar \frac{\partial q_i(\hbar, {\bf t})}{ \partial t_{k,\nu,l}} = \frac{\partial H_{k,\nu,l}({\bf{p}}, {\bf{q}}, {\bf t} ) }{ \partial p_i} \cr
\hbar \frac{\partial p_i(\hbar, {\bf t}) }{ \partial t_{k,\nu,l}} = - \frac{\partial H_{k,\nu,l}({\bf{p}}, {\bf{q}}, {\bf t} ) }{ \partial q_i} \cr
\end{array}
\right.
\eeq
It differs from equations \ref{evolution1} by the $\hbar$ factor and the fact that the Hamiltonians $ H_{k,\nu,l}({\bf{p}}, {\bf{q}}, {\bf t} )$ are non-autonomous now, i.e. they may explicitly depend on the times ${\bf t}$.

It is also interesting to remark that this rescaling implies a rescaling of the quadratic differential
$$
\left[\Tr (L(x)^2)\right] \left(dx\right)^2 \to \hbar^2 \left[\Tr (L(x)^2)\right]  \left(dx\right)^2
$$
used to define the base of the corresponding Hitchin integrable system.

\bigskip

Let us now study how the differential equation in $x$ gets deformed by introducing $\hbar$. This depends on the rank of $L_{0,r_0}$. We use the representative of the orbit considered in remark 3.7 of section \ref{SL2Action}.

\begin{itemize}
\item Let us first consider the case $L_{0,r_0} = \sigma_3$. After rescaling, the differential equation in $x$ reads
\bea \label{hintroduction}
\hbar^{d_x} \frac{\partial}{\partial x} \Psi(x,{\bf t},\hbar) &=& 
\Big[\hbar^{-r_0 d_x} \left(\left[L(x,{\bf t})\right]_{1,1} -\left[L(x,{\bf t}) \right]_{2,2} \right) + \hbar^{-d_x(r_0+1)} \left[L(x,{\bf t}) \right]_{1,2}\cr
&& + \hbar^{-d_x(r_0-1)} \left[L(x,{\bf t}) \right]_{2,1} \Big]\Psi(x,{\bf t},\hbar)
\eea
where we made explicit any dependence in $\hbar$. Remember that in all cases, $d_x = \frac{1}{r_0+1}$. It is natural to gauge transform\footnote{This gauge transformation acts by conjugation on the Lax matrices and does not change the compatibility conditions.} the system by
$$
\Psi(x,{\bf t},\hbar) \to \left( \begin{matrix} \hbar^{\frac{d_x}{2}} & 0 \cr 0 & \hbar^{-\frac{d_x}{2}}\cr \end{matrix} \right) \Psi(x,{\bf t},\hbar) 
$$
and multiply equation \eqref{hintroduction} by $\hbar^{r_0 d_x}$ to get the $\hbar$-deformed equation:
$$
\hbar \frac{\partial}{\partial x} \Psi(x,{\bf t},\hbar) = L(x,{\bf{p}}, {\bf{q}}, {\bf t} ) \Psi(x,{\bf t},\hbar)
$$
where the Lax matrix depends on $\hbar$ only through the solutions $({\bf{p}}, {\bf{q}})$ of Hamilton's equations.

Let us also notice that the rescaling can be expressed at the level of the spectral Darboux coordinates as
$$
\forall \,i \in\llbracket1, g\rrbracket \, : \; \left\{
\begin{array}{l}
 p_i \to \hbar^{r_0 d_x} p_i \cr
 q_i \to \hbar^{d_x} q_i \cr
 \end{array}\right. .
 $$

\item Let us now consider the case $L_{0,r_0} = \sigma_+$. The rescaled differential equation reads
\bea \label{hintroduction2}
\hbar^{d_x} \frac{\partial}{\partial x} \Psi(x,{\bf t},\hbar) &=& 
\Big[\hbar^{-\frac{2 r_0-1}{2} d_x} \left(\left[L(x,{\bf t})\right]_{1,1} -\left[L(x,{\bf t}) \right]_{2,2} \right) + \hbar^{- r_0 d_x} \left[L(x,{\bf t}) \right]_{1,2}\cr
&& + \hbar^{-d_x(r_0-1)} \left[L(x,{\bf t}) \right]_{2,1} \Big]\Psi(x,{\bf t},\hbar)
\eea
where $d_x =\frac{2}{2r_0+1}$. Performing the gauge transformation
$$
\Psi(x,{\bf t},\hbar) \to \left( \begin{matrix} \hbar^{\frac{d_x}{4}} & 0 \cr 0 & \hbar^{-\frac{d_x}{4}}\cr \end{matrix} \right) \Psi(x,{\bf t},\hbar) 
$$
one gets from \eqref{hintroduction2} the equation
$$
\hbar \frac{\partial}{\partial x} \Psi(x,{\bf t},\hbar) = L(x,{\bf{p}}, {\bf{q}}, {\bf t} ) \Psi(x,{\bf t},\hbar)
$$
which is the same as above.

The spectral Darboux coordinates are rescaled according to
$$
\forall \,i \in\llbracket1, g\rrbracket \, : \; \left\{
\begin{array}{l}
 p_i \to \hbar^{\frac{2r_0-1}{2r_0 +1}} p_i \cr
 q_i \to \hbar^{\frac{2}{2r_0+1}} q_i \cr
 \end{array}\right. .
 $$

\end{itemize}

\section{Topological recursion for isomonodromic systems and quantum curve\label{TRSection}}

For any $\mathfrak{g}=\mathfrak{sl}_2(\mathbb{C})$ valued  rational function $L(x)$, we have explained how to build an isomonodromic system
$$
\left\{
\begin{array}{l}
\frac{\partial }{\partial x} \Psi(x,t) = L(x,t) \Psi(x,t)\cr
\frac{\partial }{\partial t} \Psi(x,t) = A(x,t) \Psi(x,t) \cr
\end{array}
\right.
$$
such that $L(x,t=0) = L(x)$ and the compatibility condition
$$
\frac{\partial L}{\partial t}-\frac{\partial A}{\partial x}= [A,L]
$$
has a non-autonomous Hamiltonian formulation.

Moreover, we could introduce a parameter $\hbar$ in order to replace the above system by
\beq\label{LaxPair}
\left\{
\begin{array}{l}
\hbar \frac{\partial }{\partial x}  \Psi(x,t,\hbar) = L(x,t,\hbar) \Psi(x,t,\hbar)\cr
\hbar \frac{\partial }{\partial t}  \Psi(x,t,\hbar) = A(x,t,\hbar) \Psi(x,t,\hbar) \cr
\end{array}
\right.
\eeq
with the compatibility condition
\beq\label{eq-compatibility-hbar}
\hbar \frac{\partial L(x,t,\hbar)}{\partial t} - \hbar\frac{\partial  A (x,t,\hbar)}{\partial x} = [A(x,t,\hbar),L(x,t,\hbar)]
\eeq
where $\underset{{\hbar \to 0}}{\lim} L(x,t,\hbar) \in \hat{\mathfrak{g}}$ is non-vanishing.

We can now state the main result of this article.

\bt\label{main-theorem}
The $\hbar$-dependent isomonodromic systems built in section \ref{section-introduction-h} satisfy the topological type property with $\om_{0,2}$ given by the unique Bergman kernel on the classical spectral curve obtained as the compactification of $\Sigma^{cl}:=\left\{(x,y) \in \overline{\mathbb{C}}^{\,2} | \underset{\hbar\to 0}{\lim} \det \left(y -L(x,t,\hbar)\right) = 0 \right\}$.
\et

\proof{
The proof is technical and consists in checking one by one the conditions of the topological type property. This is done in the various sections of appendix \ref{AppendixA}.
}

\medskip

As a corollary, this implies that the corresponding  isomonodromic tau function can be computed by topological recursion.
\begin{corollary}
The tau function $\tau_{BM}(\mathbf{t},\hbar)$ of the $\hbar$-dependent isomonodromic systems of section \ref{section-introduction-h} reads
$$
\tau_{BM}(\mathbf{t},\hbar) = \alpha \exp \left[\sum_{g=0}^\infty \hbar^{2g-2} F_g(\mathbf{t}) \right]
$$
where $\alpha$ is a constant independent of $t$ and the coefficients $F_g(t)$ are computed by topological recursion. This also implies that $\tau_{BM}(\mathbf{t},\hbar)$ is a generating function of intersections of tautological classes in the Deligne-Mumford compactification of the moduli space of Riemann surfaces.
\end{corollary}
The above result can be interpreted as a generalization of Kontsevich-Witten theorem \cite{Kontsevich-KdV} by bringing a correspondence between isomonodromic tau functions and generating functions of correlators of CohFT's. It also provides a set of Virasoro constraints annihilating this tau function.

\medskip

In \cite{Reconstruction}, it was proved that one can construct a differential equation 
$$
\hbar \frac{d}{dx} \Psi(x,\hbar) = L(x,\hbar) \Psi(x,\hbar)
$$
with the topological type property whenever the Newton polytope of the classical spectral curve $P(x,y) = \underset{\hbar \to 0}{\lim} \det (y-L(x,\hbar)) = 0$ does not have any interior point. In particular, this procedure consists in associating a linear differential operator $\hbar \frac{d}{dx} - L(x,\hbar)$ to a classical spectral curve $P(x,y) = 0$. In the literature, this linear differential operator is often referred to as a ``quantum curve'' associated to a classical spectral curve $P(x,y)=0$. Up to now, the only examples of quantum curves obtained from a classical spectral curve of genus $0$ with interior points in its Newton polytope were given by the Painlev\'e equations \cite{MarchalIwaki}.

However, theorem \ref{main-theorem} can be used to provide a map that associates a quantum curve to any genus $0$ hyperelliptic classical spectral curve. Indeed, let us consider a genus $0$ hyperelliptic classical spectral curve presented in the form
$$
y^2 = Q(x)
$$
where $Q(x)$ is a rational function. From the pole structure of $Q(x)$ we may present this equation as the characteristic polynomial of a $\mathfrak{sl}_2$ valued function $L(x)$ with fixed poles. One can then apply the result of the present work to define a $\hbar$-deformation of $L(x)$ so that one gets a differential operator $\hbar \frac{d}{dx} - L(x,\hbar)$ satisfying the topological type property, hence producing a quantum curve whose symbol in the $\hbar \to 0$ limit is precisely $y^2-Q(x) = 0$.

It is worth noticing that, in this setup, the spectral curve of $L(x,\hbar)$ has non-vanishing genus for $\hbar \neq 0$ and the classical spectral curve corresponds to a fully degenerate curve in this family.

\section{Examples \label{SectionExamples}}
\subsection{Painlev\'e examples}
Let us apply the formalism to the case of Painlev\'e equations, finding the Hamiltonians and the auxiliary systems. All these cases correspond to $\mathfrak{sl}_2(\mathbb{C})$ cases with $r= 2$ giving rise to $D=4$ and a reduced space of dimension 2. 

\subsubsection{Painlev\'e I \label{P1Section}}
Let us discuss in details the example leading to the Painlev\'e I equation, showing how our procedure produces the deformation considered in \cite{MarchalIwaki}. For this purpose, let us consider 
$$
\hat{\mathfrak{g}} := \left\{L(x) := \sum_{i=0}^{2} L_{0,i} x^i  \, , \, (L_{0,i})_{0\leq i\leq 2} \in \mathfrak{sl}_2(\mathbb{C})^{3} \right\}/ SL_2
$$
with
$$
L_{0,2} = \left( \begin{matrix}
0 & 1 \cr
0 & 0 \cr
\end{matrix}
\right) 
$$
 fixed. This is an example where ,
$n=0$, $r_0 = 2$
and $L_{0,r_0}$  does not have rank 2. Thus, the characteristic polynomial of $L(x)$ has degree 3 in $x$.

The non-Casimir Hamiltonians are given by 
$$
H_{2,0,0} = \Res_{x \to \infty} x^{-1} \Tr \left[ L(x) \right]^2 \, dx \qquad \hbox{and} \qquad
H_{2,0,1} = \Res_{x \to \infty}  x^{-2} \Tr \left[ L(x) \right]^2 \, dx 
$$
with the associated auxiliary matrices
$$
\frac{A_{2,0,0}}{2} =  \left[x^{-1}  L(x) \right]_+ =  \left( \begin{matrix}
0 & 1 \cr
0 & 0 \cr
\end{matrix}
\right)  \, x +L_{0,1}  \qquad \hbox{and} \qquad
\frac{A_{2,0,1}}{2} =  \left[ x^{-2} L(x) \right]_+ = \left( \begin{matrix}
0 & 1 \cr
0 & 0 \cr
\end{matrix}
\right)   .
$$
Let us now explicitly express a point in a given symplectic leaf. Denoting
$$
\forall\, i\in\llbracket 1,2\rrbracket \, , \; L_{0,i} := \left( \begin{matrix}
u_i & v_i \cr
w_i & -u_i \cr
\end{matrix}
\right),
$$ 
the coefficients of the characteristic polynomial read
$$
\left\{
\begin{array}{l}
H_{2,0,0} = 2 (u_0^2 + v_0 w_0) \cr
H_{2,0,1} =  2( 2u_1 u_0 + v_1 w_0 + w_1 v_0) \cr
H_{2,0,2} =2 (w_0 +u_1^2+ v_1 w_1) \cr
H_{2,0,3} = 2 w_1 \cr
\end{array}
\right.
.
$$

One shall now restrict to a symplectic leaf by setting $H_{2,0,3} = 2 \alpha_3 \neq 0$ and $H_{2,0,2} = 2 \alpha_2$ and map it to a point in the reduced space by fixing the value of $H_{2,0,1} = 2 \alpha_1$ and $u_1 = 0$. A point in this reduced space hence reads
$$
L(x) = \left( \begin{matrix}
0 & 1 \cr
0 & 0 \cr
\end{matrix}
\right) x^2 +
\left( \begin{matrix}
0 & v_1 \cr
\alpha_3 & 0 \cr
\end{matrix}
\right) x+
\left( \begin{matrix}
u_0 & v_1^2 + \frac{\alpha_1}{\alpha_3} - v_1 \frac{\alpha_2}{\alpha_3}\cr
\alpha_2 - v_1 \alpha_3 & -u_0 \cr
\end{matrix}
\right) .
$$

We can now identify the isomonodromic time $t$ with $t_{2,0,0}$ leading to
$$
L(x) = \left( \begin{matrix}
0 & 1 \cr
0 & 0 \cr
\end{matrix}
\right) x^2 +
\left( \begin{matrix}
0 & v_1 \cr
\alpha_3 & 0 \cr
\end{matrix}
\right) x+
\left( \begin{matrix}
u_0 & v_1^2 + \frac{\alpha_1}{\alpha_3} - v_1 \frac{\alpha_2}{\alpha_3} + 2 t\cr
\alpha_2 - v_1 \alpha_3 & -u_0 \cr
\end{matrix}
\right) .
$$
One indeed has
$$
\frac{\partial \tilde{L}(x,t)}{ \partial t_{2,0,0}} = \frac{\partial \tilde{A}_{2,0,0} }{\partial x}
$$
where 
$$
\frac{\tilde{A}_{2,0,0}}{2} =  \left[x^{-1}  \tilde{L}(x,t) \right]_+ = \left( \begin{matrix}
0 & 1 \cr
0 & 0 \cr
\end{matrix}
\right)  \, x + \beta 
$$
where $\beta$ is a matrix independent of $x$.

The spectral Darboux coordinates $(p,q)$ are given by
$$
\left\{
\begin{array}{l}
q = v_1-\frac{\alpha_2}{\alpha_3} \cr
p = u_0 \cr
\end{array}
\right.
$$
leading to the representation
$$
L(x)   = \left( \begin{matrix}
p & x^2+\left(q+\frac{\alpha_2}{\alpha_3}\right) x + q \left(q+\frac{\alpha_2}{\alpha_3}\right) +\frac{\alpha_1}{\alpha_3}+2 t \cr
\alpha_3 (x-q) & -p \cr
\end{matrix}
\right) .
$$
One recovers  the Lax representation of Painlev\'e I used in \cite{MarchalIwaki} by setting $\alpha_1 = \alpha_2 = 0$ and $\alpha_3=  1$. The Hamiltonian of the system is given by $H_{2,0,0}$.

Hence, our isomonodromic system takes the form
$$
\left\{
\begin{array}{l}
\frac{\partial}{ \partial x} \Psi = \left[ \left( \begin{matrix}
0 & 1 \cr
0 & 0 \cr
\end{matrix}
\right)  x^2 + L_{0,1} x +L_{0,0} + 2 t_{2,0,0}  \left( \begin{matrix}
0 & 1 \cr
0 & 0 \cr
\end{matrix}
\right) \right] \Psi \cr
\frac{\partial}{\partial t_{2,0,0}} \Psi = \left[ \left( \begin{matrix}
0 & 1 \cr
0 & 0 \cr
\end{matrix}
\right)  \, x + \beta  \right] \Psi  \cr
\end{array}
\right. .
$$

We shall now detail the introduction of the parameter $\hbar$. From the computation of 
$$
\Tr(\tilde{L}(x,t)^2) = x^3 H_{2,0,3} + x^2 H_{2,0,2} + x H_{2,0,1} + H_{2,0,0}
$$
one can read the homogeneity degree of the Hamiltonians with respect to $x$. If $x$ has degree $d_x$ and imposing $H_{2,0,3}$ of degree 0\footnote{This assumption can be released easily but leads to more complicated expressions.}, one gets that  $H_{2,0,i}$ has degree
$$
d_{2,0,i} = (3-i)d_x.
$$

From equation \eqref{Ham}, one can compute the value of the Casimirs in terms of the entries of the matrix leading to
$
d_{t_{2,0,0}} = 2 d_x.
$
The condition
$
d_{t_{2,0,0}} + d_{2,0,0} = 2
$
imposes
$
d_x = \frac{2}{5}
$
as expected for the Painlev\'e I case from \cite{MarchalIwaki}.

\subsubsection{Painlev\'e II}
Let us consider again $n=0$ and $r_0 = 2$ but now fix
  $L_{0,2} =\sigma_3$. This implies that the characteristic polynomial of $L(x)$ has degree 4 in opposition to the Painlev\'e I case above.

As in the preceding section, the non-Casimir Hamiltonian are given again by 
$$
H_{2,0,0} = \Res_{x \to \infty} x^{-1} \Tr \left[ L(x) \right]^2 \, dx \qquad \hbox{and} \qquad
H_{2,0,1} = \Res_{x \to \infty}  x^{-2} \Tr \left[ L(x) \right]^2 \, dx  
$$
with the associated auxiliary matrices
$$
\frac{A_{2,0,0}}{2} =  \left[x^{-1}  L(x) \right]_+ = \sigma_3 \, x +L_{0,1} \qquad \hbox{and} \qquad
\frac{A_{2,0,1}}{2} =  \left[ x^{-2} L(x) \right]_+ = \sigma_3  .
$$

We can compute the spectral Darboux coordinates for this system. For simplicity let us consider a symplectic leaf of the form $(H_{2,0,3},H_{2,0,2},H_{2,0,1}) = (0,\alpha_2,\alpha_1)$.  Considering a representative of the reduced orbit as before, one has a  Lax matrix of the form
$$
L(x) = \sigma_3 x^2 + \left( \begin{matrix}
0 & v_1 \cr
1 & 0 \cr
\end{matrix}
\right) x + \left( \begin{matrix}
u_0 & v_0 \cr
w_0 & -u_0 \cr
\end{matrix}
\right) 
$$ 
with $2u_0+v_1 = \alpha_2$ and $v_0+v_1 w_0 = \alpha_1$. One obtains the spectral Darboux coordinates
$$
\left\{
\begin{array}{l}
q = -w_0 \cr
p = q^2+u_0 \cr
\end{array}
\right.
$$ giving
$$
L(x) = \sigma_3 x^2 + \left( \begin{matrix}
0 & \alpha_2- 2(p-q^2) \cr
1 & 0 \cr
\end{matrix}
\right) x + \left( \begin{matrix}
(p-q^2)& \alpha_1+q \left[\alpha_2-2(p-q^2)\right] \cr
-q & -(p-q^2)\cr
\end{matrix}
\right) .
$$ 
Remark that $\tilde{p} = p-q^2$ gives an alternative Darboux coordinate dual to $q$. In terms of the coordinates $(\tilde{p},q)$, one gets
$$
L(x) = \sigma_3 x^2 + \left( \begin{matrix}
0 & \alpha_2- 2\tilde{p} \cr
1 & 0 \cr
\end{matrix}
\right) x + \left( \begin{matrix}
\tilde{p}& \alpha_1+q \left[\alpha_2-2\tilde{p}\right] \cr
-q & -\tilde{p}\cr
\end{matrix}
\right) .
$$

Following the general procedure, one can identify the isomonodromic time with $t_{2,0,0}$ and consider
$$
\tilde{L}(x,t) := \sigma_3 x^2 + L_{0,1} x +L_{0,0} + 2 t  \sigma_3
$$
leading to the isomonodromic system 
$$
\left\{
\begin{array}{l}
\frac{\partial}{\partial x} \Psi = \left[ \sigma_3  x^2 + L_{0,1} x +L_{0,0} + 2 t_{2,0,0}  \sigma_3 \right] \Psi \cr
\frac{\partial}{\partial t_{2,0,0}} \Psi = \left[ \sigma_3 \, x + \beta  \right] \Psi  \cr
\end{array}
\right. .
$$
The explicit expression of a point in the reduced space can be performed as before and allows to introduce the parameter $\hbar$ in the same way. One finds the homogeneity degrees
$$
d_{2,0,0} = 4 d_x
\qquad \hbox{and} \qquad
d_{t_{2,0,0}} = 2 d_x
.
$$
This implies that the rescaling should follow from
$$
d_x = \frac{2}{4+2} = \frac{1}{3} .
$$

A Lax representation of Painlev\'e II is recovered by setting $H_{2,0,1} = 0$, $H_{2,0,2} = t$, $H_{2,0,3} = \theta$. The Hamiltonian of the system is given by $H_{2,0,0}$.

\subsubsection{Painlev\'e III}
In order to recover Painlev\'e III equation, let us consider the case where $n=1$, $r_0 =0$, $a_1 = 0$, $r_1=2$ and fix $L_0 = \sigma_3$ to get an element of the form
$$
L(x) = \sigma_3 +L_1 x^{-1} +L_2 x^{-2} .
$$
The non-Casimir Hamiltonians are
$$
H_{2,1,1} = - \Res_{x \to 0} \Tr \left[ L(x) \right]^2 \, dx  \qquad \hbox{and} \qquad
H_{2,1,2} = - \Res_{x \to 0}  x \Tr \left[ L(x) \right]^2 \, dx 
$$
with the associated auxiliary matrices
$$
\frac{A_{2,1,1}}{2} =  \left[  L(x) \right]_+ = - L_0 \qquad \hbox{and} \qquad
\frac{A_{2,1,2}}{2} =  \left[ x L(x) \right]_+ = - L_0 x - L_1.
$$
The Hamiltonian of the system is $H_{2,1,2}$ while $H_{2,1,1}$ should be fixed by symplectic reduction.

In order to recover a known representation of Painlev\'e III, one should consider some other auxiliary matrices by shifting $A_{2,1,2}$ by a matrix proportional to $L(x)$. We shall then recover the Lax representation considered in \cite{LoopAlgebraHarnad}. For this purpose, one defines the auxiliary matrix
$$
A(x) := {A_{2,1,2} } + 2 x L(x) =   \left[ xL_0 +L_1  -L_2 x^{-1} \right]
$$
obtained from \eqref{Psigma} with $\sigma = 2$. Hence
$$
\frac{\partial A(x)}{ \partial x}  =  \frac{1}{2} \left[ L_0   +L_2 x^{-2} \right] .
$$

This motivates the introduction of a time $t$ through a rescaling $L_0 \to t L_0$ and $L_2 \to t L_2$ leading to the isomonodromic system
$$
\left\{
\begin{array}{l}
\frac{\partial}{\partial x} \Psi = \left[ t \sigma_3  + L_{1} x^{-1} +t L_{2} x^{-2} \right] \Psi \cr
\frac{\partial}{\partial t} \Psi = \left[ x t \sigma_3 +L_1  - t L_2 x^{-1} \right]\Psi  \cr
\end{array}
\right. .
$$

Let us now introduce $\hbar$. We cannot apply the general result of the preceding section since we have considered $\sigma \neq 0$. 
However, we can take $d_x=0$ which immediately imposes that $t$ should be of degree 1 and thus $H$ as well and we recover the deformation of \cite{MarchalIwaki}.

\subsubsection{Painlev\'e IV}
Let us consider the case where $n=1$, $r_0 =1$, $r_0=1$ and $L_{0,1} = \sigma_3$,
$$
L(x) = L_{0,1} x +L_{0,0} +\frac{L_{1,1}}{ x-a_1}
$$
The non-Casimir invariant functions are
$$
H_{2,0,0} = \underset{x \to \infty}{ \Res} x^{-1} \Tr \left[ L(x) \right]^2 \, dx \qquad \hbox{and} \qquad 
H_{2,1,1} = - \underset{x \to a_1}{ \Res}  \Tr \left[ L(x) \right]^2 \, dx  
$$
with the associated auxiliary matrices
$$
\frac{A_{2,0,0}}{2} =  \left[  x^{-1} L(x) \right]_+ = L_{0,1} \qquad \hbox{and} \qquad 
\frac{A_{2,1,1}}{2} =  - \left[ L(x) \right]_- = - L_{1,1} (x-a_1)^{-1} .
$$
The Hamiltonian is $H_{2,1,1}$ while $H_{2,0,0}$ should be fixed by symplectic reduction.

Following the general procedure for $n \geq 1$, the isomonodromic time $t_{2,1,1}$ should be identified with the position $a_1$ of the second pole. This leads to the isomonodromic system
$$
\left\{
\begin{array}{l}
\frac{\partial}{\partial x} \Psi = \left[ L_{0,1} x +L_{0,0} +\frac{L_{1,1}}{ x-t_{2,1,1}} \right] \Psi \cr
\frac{\partial}{\partial t_{2,1,1}} \Psi = - \frac{L_{1,1}}{ x-t_{2,1,1}}   \Psi  \cr
\end{array}
\right. .
$$
The introduction of $\hbar$ follows from the general procedure with $d_{2,1,1} = 3$ leading to 
$$
d_x = \frac{1}{2} .
$$

\subsubsection{Painlev\'e V}
Let us consider the case where $n=2$, $r_0 =0$,  $r_1=1$,  $r_2 = 1$, $a_1 = 0$ and $a_2 = 1$,
$$
L(x) = L_{0,0}  +L_{1,1} x^{-1} +\frac{L_{2,1}}{ x-1} 
$$
with $L_{0,0} = \sigma_3$.
We consider the non-Casimir functions
$$
H  =  - \Res_{x \to \infty} x \Tr \left[ L(x) \right]^2 \, dx  \qquad \hbox{and} \qquad
a = \Res_{x \to \infty}  \Tr \left[ L(x) \right]^2 \, dx  
$$
with the associated auxiliary matrices
$$
\frac{A}{2} = -  \left[  x L(x) \right]_+ = - x L_{0,0}  - L_{1,1} \qquad \hbox{and} \qquad 
\frac{A'}{2} =  \left[ L(x) \right]_+ = L_{0,0} .
$$
In order to recover Painlev\'e V equation, following \cite{RMatrixHarnad,ItsProkhorov}, one shall consider the Hamiltonian
$$
H^*:= H -\frac{a^2}{4} .
$$
One recovers a Lax representation of Painlev\'e V by considering $\sigma  = 1$ in \eqref{Psigma} leading to the auxiliary matrix
$$
A_1 = \left[x L(x) + a L(x)\right]_+ .
$$
As in Painlev\'e III case, one obtains an isomonodromic system by rescaling $L_{0,0} \to t L_{0,0}$ and $H^* \to \frac{H^*}{t}$.

One shall now introduce $\hbar$. For this purpose, one can see that $H^*$ has the same degree as $t^2$ if $x$ is independent of $\hbar$ leading to the rescaling described in \cite{MarchalIwaki}.

\subsubsection{Painlev\'e VI}
Let us consider the case where $n=3$, $r_0 =-1$,  $a_1 = 0$, $a_2 = 1$ and $r_1 = r_2 =r_3 = 1$, i.e. the simplest  Fuchsian system.

As in any Fuchsian system, one finds an isomonodromic system by identifying $t = a_3$ (or any of the simple poles) and considering the Hamiltonian
$H_{2,3,1} $.

From the general procedure to introduce $\hbar$, one has that $x$ and $t$ have the same homogeneity degree. We present below the case of general Fuchsian systems that includes Painlev\'{e} VI.

\subsection{Fuchsian systems}
For $N=2$, one can consider a Fuchsian system with an arbitrary number of poles $n \geq 3$ such that
$$
L(x) = \sum_{i=1}^n \frac{L_{i,1}}{x-a_i} .
$$
We shall consider here the case where one fixes  $L_{0,-1} = - \underset{i=1}{\overset{n}{\sum}} L_{i,1} = \sigma_3$, other cases being easily obtained in a similar way.
One can assume that $a_1=0$ and $a_2 = 1$. The remaining $(a_j)_{3\leq j\leq n}$ can be identified with isomonodromic times: $t_i=a_{i+2}$, associated with the Hamiltonians $H_i:=H_{2,i+2,1}$ with $i\in\llbracket 1,n-2\rrbracket$.

Let us denote 
$$
L_{i,1} = u_i \sigma_3 + v_i \sigma_+ + w_i \sigma_- .
$$
One now introduces $\hbar$ by rescaling
$$
u_i \to \hbar u_i \qquad \hbox{and} \qquad v_i \to \hbar^2 v_i
$$ 
while leaving all other quantities invariant. This implies
$$ 
H_i \to \hbar^2 H_i
\qquad \hbox{and} \qquad 
t_i \to t_i .
$$
Since we are in a case where $r_0 = -1$ which was not explicitly detailed in the introduction of $\hbar$, let us use this opportunity to study it more carefully.

The full isomonodromic system reads
$$
\left\{
\begin{array}{l}
\frac{d}{dx} \Psi = \left[\underset{i=1}{\overset{n}{\sum}} \frac{\hbar^{-1} u_i \sigma_3 + \hbar^{-2} v_i \sigma_+ + w_i \sigma_- }{x-a_i} \right] \Psi \cr
\frac{d}{da_{i+2}} \Psi = \left[\frac{\hbar^{-1} u_i \sigma_3 + \hbar^{-2} v_i \sigma_+ + w_i \sigma_-}{x-a_i} + \hbar^{-1} c_i \sigma_3\right] \Psi \cr
\end{array}
\right. 
$$
where $c_i$ is a constant independent of $\hbar$. After multiplying by $\hbar$ and performing the gauge transformation
$$
\Psi \to \left( \begin{matrix} 
\hbar^\frac{1}{2} & 0 \cr
0 & \hbar^{-\frac{1}{2}} \cr
\end{matrix}
\right) \Psi,
$$
one gets the system
$$
\left\{
\begin{array}{l}
\hbar \frac{d}{dx} \Psi = \left[\underset{i=1}{\overset{n}{\sum}}  \frac{ u_i \sigma_3 + v_i \sigma_+ + w_i \sigma_- }{x-a_i} \right] \Psi \cr
\hbar \frac{d}{da_{i+2}} \Psi = \left[\frac{  u_i \sigma_3 +  v_i \sigma_+ + w_i \sigma_-}{x-a_i} + c_i \sigma_3\right] \Psi \cr
\end{array}
\right. . 
$$

\subsection{Second elements of the Painlev\'e hierarchies}
Let us consider some simple higher dimensional examples by remaining in the $\mathfrak{g}=\mathfrak{sl}_2(\mathbb{C})$ case. We consider examples leading to the second elements of some of the Painlev\'e hierarchies.

\subsubsection{Painlev\'e $\text{II}_2$ hierarchy}
Let us first consider $n=0$, $r_0 = 3$ and $L_{0,3} =\sigma_3$, 
$$
L(x) := \sum_{i=0}^{3} L_{0,i} x^i .
$$
One considers the non-Casimir invariant functions 
$$
H_{2,0,0} = \Res_{x \to \infty} x^{-1} \Tr L(x)^2 dx \quad , \quad 
H_{2,0,1} = \Res_{x \to \infty} x^{-2} \Tr L(x)^2 dx \quad \hbox{and} \quad
H_{2,0,2} = \Res_{x \to \infty} x^{-3} \Tr L(x)^2 dx 
$$
together with the auxiliary matrices
\beaa
\frac{A_{2,0,0}}{2} &=&  \left[x^{-1} L(x) \right]_+  = \sigma_3 x^2 + L_{0,2} x + L_{0,1}, \cr
\frac{A_{2,0,1}}{2} &=&  \left[x^{-2}  L(x) \right]_+  = \sigma_3 x + L_{0,2},\cr
\frac{A_{2,0,2}}{2} &=&  \left[x^{-3} L(x) \right]_+  = \sigma_3.
\eeaa
One shall proceed by symplectic reduction by fixing the value of $H_{2,0,2}$.

One can make it an isomonodromic system by considering\footnote{For completeness, we introduce here the  two isomonodromic times corresponding to this four dimensional symplectic space even though we only need one for applying our procedure.}
$$
\tilde{L}(x) := L(x) + 2 t_{2,0,1} \sigma_3 + 2 t_{2,0,0} \left[ 2 \sigma_3 x + L_{0,2}\right] .
$$
The isomonodromic systems is a Lax representation of $(\text{P II}_2)_2$ \cite{Chiba,Chiba2}.

To introduce $\hbar$, one sees that $H_{2,0,0}$ (resp. $H_{2,0,1}$) has homogenous degree $d_{2,0,0} = 6 d_x$ (resp. $d_{2,0,1} = 5 d_x$) with respect to $x$ while $d_{t_{2,0,0}} = 2 d_x$ and $d_{t_{2,0,1}} = 3 d_x$. Hence, one shall consider
$$
d_x = \frac{2 }{2+6} = \frac{2}{ 3+5} = \frac{1}{4}
$$
for the rescaling.

\subsubsection{Painlev\'e IV hierarchy}
Let us consider $n=1$, $r_0 = 2$ and $r_1= 1$ together with  $L_{0,2} =\sigma_3$, 
$$
L(x) := \sum_{i=0}^{2} L_{0,i} x^i + L_{1,1} x^{-1} .
$$

One considers the non-Casimir invariant functions 
$$
H_{2,0,0} = \Res_{x \to \infty} x^{-1} \Tr L(x)^2 dx \quad , \quad
H_{2,0,1} = \Res_{x \to \infty} x^{-2} \Tr L(x)^2 dx \quad \hbox{and} \quad
H_{2,1,1} = \Res_{x \to \infty}  \Tr L(x)^2 dx 
$$
together with the auxiliary matrices
\beaa
\frac{A_{2,0,0}}{2} &=&  \left[x^{-1} L(x) \right]_+  = \sigma_3 x + L_{0,1} ,\cr
\frac{A_{2,0,1}}{2} &=&  \left[x^{-2}  L(x) \right]_+  = \sigma_3 ,\cr
\frac{A_{2,1,1}}{2} &=&  \left[ L(x) \right]_+  = \sigma_3 x^2 + L_{0,1}x + L_{0,0} .
\eeaa
One shall fix the value of $H_{2,0,1}$ for obtaining the reduced space.

One can make it an isomonodromic system by considering 
$$
\tilde{L}(x) := L(x) + 2 t_{2,0,0} \sigma_3 + 2 t_{2,1,1} \left[ 2 \sigma_3 x + L_{0,1}\right] +2  t_{2,1,1}^2 \sigma_3
$$
The isomonodromic systems is a Lax representation of $(\text{P IV})_2$. The main isomonodromic time is $t_{2,0,0}$ in this case.

\br
The appearance of the $t_{2,1,1}^2$ term comes from the fact that $d A_{2,1,1}/ dx$ does depend on $t_{2,1,1}$ linearly. This order 2 term appears for correcting this. In general, one could have higher order terms used for correcting such issues in higher degree examples (such as higher order elements in this hierarchy).

Once again, one only needs the introduction of the isomonodromic deformation $t_{2,0,0}$ for our purpose but we introduce the other one for making the connection with the integrable hierarchy.
\er

To introduce $\hbar$, one sees that $H_{2,0,0}$ (resp. $H_{2,1,1}$) has homogenous degree $d_{2,0,0} = 4 d_x$ (resp. $d_{2,1,1} = 5 d_x$) with respect to $x$ while $d_{t_{2,0,0}} = 2 d_x$ and $d_{t_{2,1,1}} = d_x$. Hence, one shall consider
$$
d_x = \frac{2}{2+4} = \frac{2}{1+5} = \frac{1}{3}
$$
for the rescaling.

\subsubsection{Painlev\'e I hierarchy}
Let us consider $n=0$, $r_0 = 3$ and $L_{0,3} =\sigma_+$,
$$
L(x) := \sum_{i=0}^{3} L_{0,i} x^i.
$$
One considers the non-Casimir invariant functions 
$$
H_{2,0,0} = \Res_{x \to \infty} x^{-1} \Tr L(x)^2 dx \quad , \quad 
H_{2,0,1} = \Res_{x \to \infty} x^{-2} \Tr L(x)^2 dx \quad \hbox{and} \quad
H_{2,0,2} = \Res_{x \to \infty} x^{-3} \Tr L(x)^2 dx 
$$
together with the auxiliary matrices
\beaa
\frac{A_{2,0,0}}{2} &=&  \left[x^{-1} L(x) \right]_+  = \sigma_+ x^2 + L_{0,2} x + L_{0,1} ,\cr
\frac{A_{2,0,1}}{2} &=&  \left[x^{-2}  L(x) \right]_+  = \sigma_+ x + L_{0,2} , \cr
\frac{A_{2,0,2}}{2} &=&  \left[x^{-3} L(x) \right]_+  = \sigma_+.
\eeaa
One shall fix the value of $H_{2,0,2}$ for obtaining the reduced space.

One can make it an isomonodromic system by considering 
$$
\tilde{L}(x) := L(x) + 2 t_{2,0,1} \sigma_+ + 2 t_{2,0,0} \left[ 2 \sigma_+ x + L_{0,2}\right] .
$$
The isomonodromic systems is a Lax representation of $(\text{P I})_2$.

To introduce $\hbar$, one sees that $H_{2,0,0}$ (resp. $H_{2,0,1}$) has homogenous degree $d_{2,0,0} = 5$ (resp. $d_{2,0,1} = 4 d_x$) with respect to $x$ while $d_{t_{2,0,0}} = 2 d_x$ and $d_{t_{2,0,1}} = 3 d_x$. Hence, one shall consider
$$
d_x = \frac{2}{2+5} = \frac{2}{3+4} = \frac{2}{7}
$$
for the rescaling.

\subsubsection{Painlev\'e $\text{II}_1$ hierarchy}
Let us now take $n=2$, $r_0 = 2$, $r_1=1$ and fix $L_{0,2} =\sigma_+$,
$$
L(x) := \sum_{i=0}^{2} L_{0,i} x^i + L_{1,1} x^{-1} .
$$
One considers the non-Casimir invariant functions 
$$
H_{2,0,0} = \Res_{x \to \infty} x^{-1} \Tr L(x)^2 dx \quad , \quad
H_{2,0,1} = \Res_{x \to \infty} x^{-2} \Tr L(x)^2 dx \quad \hbox{and} \quad
H_{2,1,1} = \Res_{x \to \infty}  \Tr L(x)^2 dx 
$$
together with the auxiliary matrices
\beaa
\frac{A_{2,0,0}}{2} &=&  \left[x^{-1} L(x) \right]_+  = \sigma_+ x + L_{0,1} ,\cr
\frac{A_{2,0,1}}{2} &=&  \left[x^{-2}  L(x) \right]_+  = \sigma_+ ,\cr 
\frac{A_{2,1,1}}{2} &=&  \left[ L(x) \right]_+  = \sigma_+ x^2 + L_{0,1}x + L_{0,0} .
\eeaa
One shall fix the value of $H_{2,0,1}$ for defining the reduced space.

One can make it an isomonodromic system by considering 
$$
\tilde{L}(x) := L(x) + 2 t_{2,0,0} \sigma_+ + 2 t_{2,1,1} \left[ 2 \sigma_+ x + L_{0,1}\right] +2  t_{2,1,1}^2 \sigma_+ 
$$
which gives a Lax representation of $(\text{P II}_1)_2$. The main isomonodromic time is $t_{2,0,0}$ in this case.

To introduce $\hbar$, one sees that $H_{2,0,0}$ (resp. $H_{2,1,1}$) has homogenous degree $d_{2,0,0} = 3 d_x$ (resp. $d_{2,1,1} = 4 d_x$) with respect to $x$ while $d_{t_{2,0,0}} = 2 d_x$ and $d_{t_{2,1,1}} = d_x$. Hence, one shall consider
$$
d_x = \frac{2}{2+3} = \frac{2}{1+4} = \frac{2}{5}
$$
for the rescaling.

\subsection{Polynomial of arbitrary order}
Let us show how our procedure applies to arbitrary degree polynomials as the ones coming from integrables hierarchies. We shall consider the case $n=0$ and arbitrary values of $r_0$ with $L_{0,r_0} = \sigma_3$ allowing to treat elements of the Painlev\'e $\text{II}_2$ hierarchy
$$
L(x) = \sigma_3 x^{r_0} + \sum_{i=0}^{r_0-1} L_{0,i} x^i .
$$
Other values of $L_{0,r_0}$ or values of $n$ can be treated in a similar way. The functions
$$
H_{2,0,i} := \Res_{x \to \infty} x^{-i-1} \Tr \left(L(x)^2\right) dx
$$
provide Casimir functions for $i \geq r_0$ and non-Casimir for $0 \leq i \leq r_0-1$ with associated auxiliary functions
$$
\forall \,0 \leq i \leq r_0-1 \, , \; \frac{ A_{2,0,i}}{ 2 } = \left[x^{-i-1} L(x) \right]_+ .
$$
We shall fix $H_{2,0,r_0-1}$ for defining the reduced space which has dimension $2 r_0-1 $ and is equipped with the Hamiltonians $H_{2,0,i}$ for $i\in \llbracket 0, r_0-2\rrbracket$.

We can now define the associated isomonodromic system by introducing the times $\mathbf{t} = (t_{2,0,i})_{i=0}^{r_0-2}$ by defining 
$$
\tilde{L}(x,\mathbf{t}) = \sum_{k=0}^{r_0} \tilde{L}_{0,k}(\mathbf{t}) x^k
$$
where $\tilde{L}_{0,k}(\mathbf{t})$, $k=0,\dots,r_0$, are obtained as the unique solution of the set of equations
$$
\forall \,0\leq k \leq r_0 \, , \; \forall \,0 \leq k+i+2 \leq r_0 \, , \; \frac{\partial \tilde{L}_{0,k}(\mathbf{t}) }{ \partial t_{2,0,i}} = 2 (k+1) \tilde{L}_{0,k+i+2}
$$
with initial condition
$$
\tilde{L}_{0,k}(0) = L_{0,k} .
$$

One can now introduce $\hbar$ following the general procedure leading to
$$
d_x = \frac{1}{r_0+1}
$$

\section{Conclusion and outlook \label{SectionConclusion}}

This article explains how the topological recursion can be used to express a tau function of isomonodromic deformations of $SL_2$ connections over the Riemann sphere in some particular regimes. It is natural to ask whether this procedure can be generalized to any meromorphic $G$-connection over an arbitrary Riemann surface for a reductive group $G$. It seems very likely to be the case and we plan to address this problem in a future work. From the topological recursion side, the TT properties have been defined for such a general setup in \cite{LoopLie}. On the other hand, the Hitchin integrable system is well known to be described by a similar $R$-matrix construction. It seems also possible to deform it to an isomonodromic system by de-autonomization and introduce a parameter $\hbar$ through the same steps as the ones followed in the present article. The question is to know whether we can do it in a way which gives rise to the TT properties. If the properties 2, 3, 5 and 6 of Definition \ref{DefTTproperty} should be easy to obtain, the genus $0$ property as well as the pole property will certainly prove to be harder to get, even though the $\hbar \to 0$ limit of Lax equations are likely to allow to obtain property 1, we cannot rely on a simple explicit matrix representation of the corresponding Lie algebra.

On the other hand, it should be noted that recent progress have been made for classifying such isomonodromic systems by establishing isomorphisms between moduli spaces of connections and Nakajima quiver varieties \cite{Hiroe,Woodhouse}. This allows to associate a quiver to an isomonodromic system of the type studied here. These quivers have symmetries that preserve the differential of the associated isomonodromic tau function, hence the free energies defined in our paper. In the simplest case, such symmetries give rise to the well-known Harnad duality \cite{Harnad-dual}. In particular, it means that even if our result applies only to rank 2 systems, they also imply that the tau function of Harnad's dual system can also be computed by topological recursion, giving access to higher rank systems. It would be very nice to understand better the action of the symmetries acting on quivers in terms of the topological recursion. It is highly plausible that it would shed some light on the mysterious symplectic invariance property of the partition functions computed by topological recursion.

Another natural generalization of our work would be to work with classical spectral curves embedded into $\mathbb{C}^*\times \mathbb{C}^*$ instead of $\mathbb{C}^2$. This is usually obtained by considering exponentiated variables as in the study of Gromov-Witten invariants of local CY three-folds. There exists a corresponding integrable system defined by Goncharov and Kenyon \cite{GK} which considers the space of connections on graphs on a torus. This replaces the usual Hitchin system by trading differential operators by finite difference operators. We believe that one should be able to apply our procedure to such a system to obtain a partition function computed by the topological recursion on this new type of classical spectral curve.

Even further is the possibility to quantize the whole picture (and all the generalizations described above) by replacing the classical $R$-matrix by a quantum $R$-matrix construction. Such a quantization of isomonodromic systems gives rise to the Knizhnik-Zamolodchikov-Bernard equations \cite{Harnad-KZ,Reshetikhin}. This allows us to interpret the result in terms of sections of the associated Verlinde bundle, i.e. in terms of some conformal blocks of an associated CFT. From the topological recursion perspective, we believe that it corresponds to the so-called $\beta$-deformations taking as input a non-commutative version of the classical spectral curve.

Finally, all our computations give rise to formal series in $\hbar$, where $\hbar$ can be thought of as a local coordinate around a genus $0$ singular point in the base  of Hitchin's integrable system. It would be nice to be able to consider the expansion around any fiber of the Hitchin system. However, we do not have any hope to be able to do so without introducing some non-perturbative effects breaking the $\hbar$-expansion properties. There exists a conjecture in the literature \cite{BEInt} for such a non-perturbative generalization. This conjecture was mainly motivated by some random matrix model examples. Since isomonodromic systems are generalizations of random matrix models and admit some Riemann-Hilbert representations, we could hope to use this approach to give a more firm ground to the conjecture if not proving it.

\bigskip

\bigskip


\bigskip

\bigskip


\appendix

\numberwithin{equation}{section}

\section{Topological type property and isomonodromic systems\label{AppendixA}}
In this appendix, we prove theorem \ref{main-theorem} by checking that the conditions 1-6 of the topological type property given in definition \ref{def-TT=property} are fulfilled by systems considered in the present paper.

\subsection{Proof of condition 2 : Formal expansion in $\hbar$}

There are several ways to justify the existence of a formal expansion in $\hbar$. The first one used in \cite{Deter,BBEnew,P2,MarchalIwaki} is to consider formal WKB solutions of the system of equations \eqref{LaxPair}. In this context, it is equivalent to assume that 
\beq\label{WKB} \Psi(x,t,\hbar)=\text{exp}\left(\sum_{k=-1}^\infty \Psi^{(k)}(x,t)\hbar^k\right)=\left(\sum_{k=0}^\infty \td{\Psi}^{(k)}(x,t)\hbar^k\right)\text{exp}\left(\frac{1}{\hbar}\Psi^{(-1)}(x,t)\right) .\eeq
Since $L(x,t,\hbar)=\left(\hbar\frac{d}{dx}\Psi(x,t,\hbar)\right)\Psi^{-1}(x,t,\hbar)$ and $A(x,t,\hbar)=\left(\hbar\frac{d}{dt}\Psi(x,t,\hbar)\right)\Psi^{-1}(x,t,\hbar)$, one automatically gets that $L(x,t,\hbar)$, $A(x,t,\hbar)$ and $M(x,t,\hbar)$ admits a formal series expansion in $\hbar$ of the form:
\beq \label{FormalSeriesExpansions} L(x,t,\hbar)=\sum_{k=0}^\infty L^{(k)}(x,t)\hbar^k\,\,,\,\, A(x,t,\hbar)=\sum_{k=0}^\infty A^{(k)}(x,t)\hbar^k \,\,,\,\, M(x,t,\hbar)=\sum_{k=0}^\infty M^{(k)}(x,t)\hbar^k .\eeq

However, according to the general geometric construction presented in this article, we propose an alternative way to obtain the same result. Indeed, in this paper, we introduce the parameter $\hbar$ using rescaling of parameters and therefore it is necessary to justify the existence of formal $\hbar$ expansions \eqref{FormalSeriesExpansions} within this formalism. First of all, let us mention that after rescaling, the values of the Casimirs defining the symplectic leaf as well as the value of the moment map used for the reduction are independent of $\hbar$. In our construction, the dependance in $\hbar$ only comes from the values of the Darboux coordinates $\left(p_i(\hbar),q_i(\hbar)\right)_{i=1}^g$. Given the non-autonomous Hamiltonian $H$ and the associated isomonodromic time $t$ built in section \ref{SectionLinearAuxMat}, the $\hbar$-deformed Hamilton's equations are given by:
\beq\label{Hamiltonh} 
\forall \,i\in\llbracket 1,g\rrbracket \, , \; \hbar \frac{\partial q_i(\hbar)}{\partial t}=\frac{\partial}{\partial p_i} H(\textbf{p},\textbf{q},\textbf{t})\, \text{ and }\, -\hbar \frac{\partial p_i(\hbar)}{\partial t}=\frac{\partial}{\partial q_i} H(\textbf{p},\textbf{q},\textbf{t}).\eeq
We get that $\left(p_i(\hbar),q_i(\hbar)\right)_{i=1}^g$ admit a formal series expansion in $\hbar$.\footnote{Standard analytic results prove that they admit a Taylor expansion at any order thus determining the coefficients of the formal expansion. However, the convergence of the full series are not guaranteed.} Moreover, for $\hbar=0$, $(q_i(\hbar=0),p_i(\hbar=0))_{i=1}^g$ are solutions of the algebraic equations $\frac{\partial}{\partial p_i} H_i(\textbf{p},\textbf{q},\textbf{t})=0$ and $\frac{\partial}{\partial q_i} H_i(\textbf{p},\textbf{q},\textbf{t})=0$. Sub-leading coefficients of the formal series expansions are then determined recursively using \eqref{Hamiltonh}. Finally, since $\left(p_i(\hbar),q_i(\hbar)\right)_{i=1}^g$ admit a formal series expansion in $\hbar$ and since the $\hbar$-dependence only comes from these Darboux coordinates, we get that $L(x,t,\hbar)$ has a formal expansion given by \eqref{FormalSeriesExpansions}. Indeed, by definition of the spectral Darboux coordinates of section \ref{DarbouxSection} and the reconstruction of the element $L(x,t)$ through Lagrange interpolation, $L(x,t)$ has components that are rational functions of $(p_i(\hbar),q_i(\hbar))_{i=1}^g$. Eventually, by definition, it implies that the auxiliary matrix $A(x,t,\hbar)$ also has a formal expansion given by \eqref{FormalSeriesExpansions}.

Since $\hbar\partial_x \Psi(x,t,\hbar)=L(x,t,\hbar)\Psi(x,t,\hbar)$ with $L(x,t,\hbar)$ having a formal series expansion in $\hbar$, it implies that $\Psi(x,t,\hbar)$ admit a formal WKB expansion given by equation \eqref{WKB}\footnote{It implies that $\hbar \log \Psi$ admits a formal series expansion in $\hbar$ which is equivalent to say that $\Psi$ admit a formal WKB expansion.}.We then conclude using its definition that $M(x,t,\hbar)$ admits a formal series expansion in $\hbar$. Note that this result is also a direct consequence of the equations $\hbar \partial_x M(x,t,\hbar)=\left[L(x,t,\hbar),M(x,t,\hbar)\right]$ and $\hbar \partial_t M(x,t,\hbar)=\left[A(x,t,\hbar),M(x,t,\hbar)\right]$ from which we may find recursively the coefficients of the formal series (see section \ref{sec-computation-M} for explicit formulas in the $\mathfrak{sl}_2(\mathbb{C})$ case). Finally, we conclude from definition \ref{DefWn} that the correlation functions admit an expansion in $\hbar$.

\begin{remark}
We stress again that these series expansions, as well as the WKB expansion for $\Psi(x,t,\hbar)$ are only considered as formal series in $\hbar$. In particular, the issue of convergence of these series remains and is well-known to be a difficult question involving Stokes sectors, Borel summability and tedious analytical considerations \cite{IwakiExactWKB,IwakiExactWKB2}. However, since our main goal is to prove the formal reconstruction of these expansions via the topological recursion (order by order in powers of $\hbar$), we may consider only formal series in $\hbar$ leaving out the issues of convergence of these series. 
\end{remark}

\subsection{Proof of condition 3 : Parity of the formal series expansion of the correlation functions}

Transposing \eqref{eq-compatibility-hbar} and \eqref{LaxPair}, one finds out that, for a given matrix $\Gamma\in G$ (independent of $x$ and $t$), we may define
$$
L_\Gamma(x,{\bf t},\hbar):= \Gamma L(x,{\bf t},\hbar)^{t\,} \Gamma^{-1} \quad , \quad A_\Gamma:= \Gamma A(x,{\bf t},\hbar)^{t\,} \Gamma^{-1} \text{ and } \Psi_\Gamma(x,{\bf t},\hbar)= \Psi(x,{\bf t},\hbar)^{t\,}\Gamma^{-1} .
$$
These matrices satisfy
$$
\left\{
\begin{array}{l}
\hbar \frac{d}{dx} \Psi_\Gamma(x,{\bf t},\hbar) = \Psi_\Gamma(x,{\bf t},\hbar)L_\Gamma(x,{\bf t},\hbar) \cr
\hbar \frac{d}{dt} \Psi_\Gamma(x,{\bf t},\hbar) =  \Psi_\Gamma(x,{\bf t},\hbar)A_\Gamma(x,{\bf t},\hbar) \cr
\end{array}
\right.
$$
and are related via the corresponding compatibility equation
$$
-\hbar\frac{dL_\Gamma}{ dt} + \hbar\frac{d A_\Gamma}{ dx} = \left[A_\Gamma ,L_\Gamma\right] ,
$$
where the relation between $A_\Gamma$ and $L_\Gamma$ is the same as the one between $A$ and $L$.

As an element of a reduced space, $L_\Gamma(x,{\bf t},\hbar)$ is given by Darboux coordinates $(p_i^\Gamma(\hbar,{\bf t}),q_i^\Gamma(\hbar,{\bf t}))_{i=1}^g$ satisfying the evolution equations
$$
\forall\, i\in\llbracket 1,g\rrbracket \, , \; \left\{
\begin{array}{l}
\hbar \frac{\partial q_i^\Gamma(\hbar, {\bf t})}{ \partial t_{k,\nu,l}} = - \frac{\partial H_{k,\nu,l}({\bf{p}^\Gamma}, {\bf{q}}^\Gamma, {\bf t} ) }{ \partial p_i^\Gamma} \cr
\hbar \frac{\partial p_i^\Gamma(\hbar, {\bf t})}{ \partial t_{k,\nu,l}} = \frac{\partial H_{k,\nu,l}({\bf{p}}^\Gamma, {\bf{q}}^\Gamma, {\bf t} ) }{\partial q_i^\Gamma} \cr
\end{array}
\right.
$$
since the transposition changes the sign of the Poisson bracket. Note that $H_{k,\nu,l}({\bf{p}}^\Gamma, {\bf{q}}^\Gamma, {\bf t} )$ are the same Hamiltonians as the one computed from $L(x,{\bf t},\hbar)$ as function of the Darboux coordinates. Hence, these evolution equations are the same as the one for the Darboux coordinates describing the evolution of $L(x,{\bf t},-\hbar)$. In addition, because they are given by the same polynomial equations, the values of $\left(p_i({\bf t},-\hbar),q_i({\bf t},-\hbar)\right)_{i=1}^g$ and $\left(p_i^\Gamma({\bf t},\hbar),q_i^\Gamma({\bf t},\hbar)\right)_{i=1}^g$ coincide  for $\hbar =0$.

Hence, for any $\Gamma$, $\left(p_i^\Gamma({\bf t},\hbar),q_i^\Gamma({\bf t},\hbar)\right)_{i=1}^g = \left(p_i({\bf t},-\hbar),q_i({\bf t},-\hbar)\right)_{i=1}^g$ and $L_\Gamma(x,{\bf t},\hbar)$  coincides with a point in the orbit of $L(x,{\bf t},-\hbar)$ meaning that
$$
\exists \,\Gamma \in SL_{L_{0,r_0}} \, , \; L_\Gamma(x,{\bf t},\hbar) = L(x,{\bf t},-\hbar).
$$

This is a sufficient condition for the parity property required for the TT property. Let us be more specific and construct explicitly this matrix $\Gamma$ for $\mathfrak{g} = \mathfrak{sl}_2$.

\subsubsection{Specific form of the $\Gamma$ matrix in the $\mathfrak{sl}_2$ case}
Depending on its rank, we can generically consider two possible leading orders $L_{0,r_0}$, namely $L_{0,r_0}=\sigma_3$ or $L_{0,r_0}=\sigma_+$. In both cases, one can obtain information about the matrix $\Gamma$ required to obtain the parity property.

Indeed, by definition, the matrix $\Gamma$ satisfies the identity $\Gamma L^{t\,}=L^\dagger \Gamma$ where $L^\dagger(x,{\bf t},\hbar) = L(x,{\bf t},-\hbar)$. Since $L_{0,r_0}$ is chosen independently of $\hbar$, $t$ and $x$, we may project the last identity at $x^{r_0}$ and get $\Gamma L_{0,r_0}^{\,\,\,\,t\,}=L_{0,r_0} \Gamma$ (because $L_{0,r_0}^\dagger=L_{0,r_0}$ since it is chosen fixed by the $SL_2$ action). 

\begin{itemize}\item \underline{Case $L_{0,r_0}=\sigma_3$}: In this case, $\Gamma L_{0,r_0}^{\,\,\,\,t\,}=L_{0,r_0} \Gamma$ is equivalent to $\Gamma_{1,2}=\Gamma_{2,1}=0$. Since $\Gamma\in G=SL_2$, we end up with
$$
\Gamma(t,\hbar)=\begin{pmatrix}\alpha(t,\hbar)&0\\0&\frac{1}{\alpha(t,\hbar)} \end{pmatrix} 
$$
where $\alpha(t,\hbar)$ is an appropriate function of $t$ and $\hbar$ to be determined. Note that this general form is in agreement with the Painlev\'{e} cases (except Painlev\'{e} I for which $L_{0,r_0}=\sigma_+$ that is studied below) studied in \cite{MarchalIwaki}. 

\item \underline{Case $L_{0,r_0}=\sigma_+$}: In this case, the $SL_2$ action also implies that $L_{0,r_0-1}=C+l(t,\hbar)\sigma_+$ where $C$ is a constant (i.e. independent of $t$, $\hbar$ and $x$) matrix. The identity $\Gamma L_{0,r_0}^{\,\,\,\,t\,}=L_{0,r_0} \Gamma$ is equivalent to $\Gamma_{2,1}=\Gamma_{1,2}$ and $\Gamma_{2,2}=0$. Since $\Gamma\in G=SL_2$ and $\Gamma$ is defined up to a global sign (that does not change the value of the determinant), we get that we may always choose $\Gamma_{1,2}=\Gamma_{2,1}=i$. In other words,
\beq \label{GammaSigmaPlus} 
\Gamma(t,\hbar)=\begin{pmatrix}\alpha(t,\hbar)&i\\i&0 \end{pmatrix}.
\eeq
Using the additional information on $L_{0,r_0-1}$ and projecting the identity $\Gamma L^{t}=L^\dagger \Gamma$ at order $x^{r_0-1}$ gives 
$$
\Gamma C^t-C\Gamma=l(t,-\hbar)\sigma_+\Gamma-l(t,\hbar)\Gamma \sigma_- .
$$
The l.h.s. and r.h.s. of the previous equation may be computed using \eqref{GammaSigmaPlus}
\beaa 
\Gamma C^t-C\Gamma&=&\begin{pmatrix}0&\alpha(t,\hbar) C_{2,1}-2i C_{1,1}\\ 2i C_{1,1}-\alpha(t,\hbar) C_{2,1} &0\end{pmatrix}\cr
l(t,-\hbar)\sigma_+\Gamma-l(t,\hbar)\Gamma \sigma_-&=&\begin{pmatrix}i(l(t,-\hbar)-l(t,\hbar))&0\\0&0\end{pmatrix} .
\eeaa
Hence, we end up with the fact that $l(t,\hbar)=l(t,-\hbar)$ which is equivalent to say that $l(t,\hbar)$ is an even function of $\hbar$ and $\alpha(t,\hbar) C_{2,1}=2i C_{1,1}$. Thus, in the case when $C_{2,1}\neq 0$ we get 
$$
\Gamma=\begin{pmatrix}2i\frac{C_{1,1}}{C_{2,1}}&i\\i &0\end{pmatrix}.
$$
Note that this general form as well as the fact that $l(t,\hbar)$ is an even function of $\hbar$ is verified in the case of Painlev\'{e} I developed in details in \cite{MarchalIwaki} where $l(t,\hbar)=q(t,\hbar)$ is an even function of $\hbar$ while $(C_{1,1},C_{2,1})=(0,4)$ and $\Gamma=\begin{pmatrix} 0&i\\ i&0\end{pmatrix}$. 

In the case $C_{2,1}=0$, we only get $C_{1,1}=0$ so that $C$ is proportional to $\sigma_+$ and thus may be absorbed in the $l(t,\hbar)$ factor in the definition of $L_{0,r_0-1}$. Thus, $L_{0,r_0-1}$ is proportional to $\sigma_+$ and in particular $\det L(x,t,\hbar)\overset{x\to \infty}{=}O(x^{2r_0-2})$, i.e. that $L(x,t,\hbar)$ is twice degenerate at infinity. In this highly degenerate cases, the missing factor $\alpha(t,\hbar)$ of $\Gamma$ should be studied specifically but the previous study still proves that $l(t,\hbar)$ is always an even function of $\hbar$. Note that this degenerate case cannot happen for $(\nu,r_0)=(0,1)$ since it would give $L(x)=(x+c)\sigma_+$ and thus $\det L(x)=0$.    
\end{itemize}

\subsection{Proof of property 1 : Genus 0 property}

For a given $\mathbf{t}$, the classical spectral curve is defined as
$$
\det(yI_2-L(x,\mathbf{t},\hbar))_{\hbar=0}=0=y^2+\det L^{(0)}(x,\mathbf{t}).
$$
Since $ \det L^{(0)}(x,\mathbf{t})$ is a rational function of $x$ with poles at $x \in\{a_0,a_1,\dots,a_n\}$, this defines a Riemann surface with possible punctures at those points. In order to prove the topological type property, we must prove that this always defines a Riemann surface of genus $0$. In the case of $\mathfrak{sl}_2(\mathbb{C})$, the situation is particularly simple. Indeed, taking order $\hbar^0$ into the compatibility equations leads to
$$
 \left[L^{(0)},A^{(0)}\right] 
=0.
$$
In other words, all leading orders in $\hbar$ of the differential systems must commute. In particular, this is equivalent to the fact that their eigenspaces coincide. Since they are elements of $\mathfrak{sl}_2(\mathbb{C})$, there are only two possible cases.
\begin{itemize}\item $0$ is an eigenvalue of $L^{(0)}$, and therefore it must be an eigenvalue of order $2$. This means that $L^{(0)}$ and $A^{(0)}$ are proportional to $\sigma_+$ (remember that we took the leading order $L_{0,r_0}$ equal to either $\sigma_3$ or $\sigma_+$). Projecting \eqref{LaxPair} at order $\hbar^0$ using  \eqref{WKB} we get $\tilde{\Psi}^{(0)}\partial_x \Psi^{(-1)}=L^{(0)}\tilde{\Psi}^{(0)}$. Since $\tilde{\Psi}^{(0)}$ is invertible ($\Psi \in G=SL_2(\mathbb{C})$) and $\Psi^{(-1)}$ is diagonal while $L^{(0)}$ is proportional to $\sigma_+$, we end up with a contradiction.

\item $L^{(0)}$ admits two opposite non-zero eigenvalues $\pm s_0$. This means that each eigenspace is of dimension $1$. Therefore, all matrices $A^{(0)}$ are also diagonal in this basis with two opposite eigenvalues (since they belong to $\mathfrak{sl}_2(\mathbb{C})$). Thus, the matrices $L^{(0)}$ and $A^{(0)}$ are proportional to each other.
\end{itemize}
In conclusion, the previous discussion leads to the fact that
$$
 \exists\, \alpha(x,\mathbf{t}) \in \mathbb{C} \,/\, A^{(0)}(x,\mathbf{t})=\alpha(x,\mathbf{t}) L^{(0)}(x,\mathbf{t})
$$
where $\alpha(x,\mathbf{t})\in \mathbb{C}^*$.

At the level of determinants, we end up with
$$
\det A^{(0)}(x,\mathbf{t}) =\alpha(x,\mathbf{t}) ^2 \det L^{(0)}(x,\mathbf{t}),
$$
so that the spectral curve is given by
$$
y^2+\frac{\det A^{(0)}(x,\mathbf{t})  }{\alpha(x,\mathbf{t}) ^2}=0.
$$

From section \ref{SectionLinearAuxMat}, we know that there exists an auxiliary matrix $A(x,\mathbf{t},\hbar)$ of the form
\beq \label{CrucialLaxSystem}
A(x,\mathbf{t},\hbar) = \frac{\hat{A}(x,\mathbf{t},\hbar) }{ \beta(x)}
\eeq
where $\beta(x) = 1$ or $\beta(x) = x-a_i$ and $ \hat{A}(x,\mathbf{t},\hbar)$ is a $\mathfrak{sl}_2$ valued degree $1$ polynomial in $x$. This leads to a spectral curve given by
$$
y^2+\frac{\det \hat{A}^{(0)}(x,\mathbf{t})  }{\beta(x)^2 \alpha(x,\mathbf{t}) ^2}=0 .
$$

Thus, we end up with the following theorem:
\begin{theorem}[Genus 0] The family of classical spectral curves $\Sigma_{\mathbf{t}}$ is of genus $0$. 
\end{theorem}

\subsection{Proof of property 4 : Pole property}
\subsubsection{Computation of $\left(M^{(k)}\right)_{k\geq0}$} \label{sec-computation-M}
In the $\mathfrak{sl}_2(\mathbb{C})$ case, the correlation functions are defined as polynomials of the matrix $M(x,\mathbf{t})=\Psi(x,\mathbf{t}) e_1 \Psi(x,\mathbf{t})^{-1}$.\footnote{Remember that even if $e_1$ does not belong to $\mathfrak{sl}_2$, we can still use it for proving the TT properties as explained in remark \ref{rem-gl-vs-sl}.
} It is a projection of rank $1$ satisfying $M^2=M$ with $\Tr M=1$. It admits a formal $\hbar$-expansion of the form
$$
M(x,\mathbf{t})=\sum_{k=0}^\infty M^{(k)}(x,\mathbf{t}) \hbar^k.
$$
It satisfies the differential system
$$
\left\{\begin{array}{l}
\hbar\frac{\partial M}{\partial x}=\left[L,M\right] \cr
\hbar\frac{\partial M}{\partial t}=\left[A,M\right]  .\cr
\end{array}
\right.
$$
In order to prove that the correlation functions admit singularities only at the branchpoints, we will prove this property for $M$. The strategy follows the one already used in \cite{MarchalIwaki} for Painlev\'{e} equations. Using $\det M^{(0)}=0$ and the time-differential equations, we get (See \cite{MarchalIwaki} for details)
\beq \label{M0} 
M^{(0)}(x,\mathbf{t})=\begin{pmatrix}\frac{1}{2}+\frac{(A^{(0)})_{1,1}(x,t)}{2\sqrt{-\det A^{(0)}(x,t)}}& \frac{(A^{(0)})_{1,2}(x,t)}{2\sqrt{-\det A^{(0)}(x,t)}}\\
\frac{(A^{(0)})_{2,1}(x,t)}{2\sqrt{-\det A^{(0)}(x,t)}}&\frac{1}{2}-\frac{(A^{(0)})_{1,1}(x,t)}{2\sqrt{-\det A^{(0)}(x,t)}}\end{pmatrix}=\frac{1}{2}I_2+\frac{1}{2\sqrt{-\det A^{(0)}(x,t)}}A^{(0)}(x,t) .
\eeq
Thus, it is obvious that $x\mapsto M^{(0)}(x,\mathbf{t})$ may only be singular at the branchpoints with only at most pole singularity there (note in particular the factor $\beta(x)$ in \eqref{CrucialLaxSystem} always simplifies in the previous formula so that $M^{(0)}(x,\mathbf{t})$ is regular at the poles of $A^{(0)}(x,\mathbf{t})$).
Next, following \cite{MarchalIwaki}, we may determine the matrices $\left(M^{(k)}\right)_{k\geq 1}$ recursively. However, we need to extend the strategy used in \cite{MarchalIwaki} in order to make it sufficient for our general case. The recursion provided in \cite{MarchalIwaki} is given by projecting the relation:
$$
\hbar\frac{\partial M}{\partial t}=\left[A,M\right] \text{ and } \det M=0
$$
at order $\hbar^k$. In \cite{MarchalIwaki}, only the entries $(1,1)$ and $(1,2)$ of the first identity were used since the entries $(2,1)$ and $(2,2)$ are not independent from them. However, this strategy generates formulas involving $\frac{1}{A_{2,1}^{(0)}}$ that may be problematic in the generic case (it was not necessary for the Lax pairs considered in \cite{MarchalIwaki} since the top right entries of the matrices $A^{(0)}$ considered there were simple) because they create fake poles at the zeros of $A^{(0)}_{1,2}$. In order to avoid such problems, one simply has to remember that the entry $(2,1)$ may also be used and provide alternative formulas. Following this strategy we obtain the over-determined system of equations for $(M^{(k)}(x,t)_{1,1},M^{(k)}(x,t)_{1,2},M^{(k)}(x,t)_{2,1})$ (for $k\geq 1$ we have $M^{(k)}(x,t)_{2,2}=-M^{(k)}(x,t)_{1,1}$ since $\Tr M(x,t,\hbar)=1$):
\beaa  
&&\begin{pmatrix} 0&-(A^{(0)})_{2,1}&(A^{(0)})_{1,2}\\
-2(A^{(0)})_{1,2}&2(A^{(0)})_{1,1}&0\\
2(A^{(0)})_{2,1}&0&-2(A^{(0)})_{1,1}\\
\frac{(A^{(0)})_{1,1}}{\sqrt{-\det A^{(0)}}}&\frac{1}{2}\frac{(A^{(0)})_{2,1}}{\sqrt{-\det A^{(0)}}}&\frac{1}{2}\frac{(A^{(0)})_{1,2}}{\sqrt{-\det A^{(0)}}}\end{pmatrix}
\begin{pmatrix} M^{(k)}(x,t)_{1,1}\\ M^{(k)}(x,t)_{1,2}\\ M^{(k)}(x,t)_{2,1}\end{pmatrix}\cr
&&=\begin{pmatrix} \partial_{t} M^{(k-1)}(x,t)_{1,1}-\underset{j=0}{\overset{k-1}{\sum}}\left[A^{(k-j)}(x,t),M^{(j)}(x,t)\right]_{1,1}\\ 
\partial_t M^{(k-1)}(x,t)_{1,2}-\underset{j=0}{\overset{k-1}{\sum}} \left[A^{(k-j)}(x,t),M^{(j)}(x,t)\right]_{1,2}\\
\partial_t M^{(k-1)}(x,t)_{2,1}-\underset{j=0}{\overset{k-1}{\sum}} \left[A^{(k-j)}(x,t),M^{(j)}(x,t)\right]_{2,1}\\
\underset{j=1}{\overset{k-1}{\sum}}\left( M^{(j)}(x,t)_{1,1}M^{(k-j)}(x,t)_{1,1}+M^{(j)}(x,t)_{1,2}M^{(k-j)}(x,t)_{2,1}\right)\end{pmatrix}\overset{\text{def}}{=}\begin{pmatrix}\text{RHS}^{(k)}_1(x,t)\\\text{RHS}^{(k)}_2(x,t)\\ \text{RHS}^{(k)}_3(x,t)\\ \text{RHS}^{(k)}_4(x,t)\end{pmatrix} . \cr
&& 
\eeaa

Thus we find the recursion
\footnotesize{\beaa
M^{(k)}(x,t)_{1,1}&=&\frac{1}{\det A^{(0)}(x,t)}\Big[\frac{1}{2}A^{(0)}(x,t)_{1,1}\text{RHS}^{(k)}_1(x,t)+\frac{1}{2}A^{(0)}(x,t)_{2,1}\text{RHS}^{(k)}_2(x,t)\cr
&&-A^{(0)}(x,t)_{1,1}\sqrt{-\det A^{(0)}(x,t)}\text{RHS}^{(k)}_4(x,t)\Big] ,\cr
M^{(k)}(x,t)_{1,2}&=&\frac{1}{\det A^{(0)}(x,t)}\Big[\frac{1}{2}A^{(0)}(x,t)_{1,2}\text{RHS}^{(k)}_1(x,t)-\frac{1}{2}A^{(0)}(x,t)_{1,1}\text{RHS}^{(k)}_2(x,t)\cr
&&-A^{(0)}(x,t)_{1,2}\sqrt{-\det A^{(0)}(x,t)}\text{RHS}^{(k)}_4(x,t)\Big] ,\cr
M^{(k)}(x,t)_{2,1}&=&\frac{1}{\det A^{(0)}(x,t)}\Big[-\frac{1}{2}A^{(0)}(x,t)_{2,1}\text{RHS}^{(k)}_1(x,t)+\frac{1}{2}A^{(0)}(x,t)_{1,1}\text{RHS}^{(k)}_3(x,t)\cr
&&-A^{(0)}(x,t)_{2,1}\sqrt{-\det A^{(0)}(x,t)}\text{RHS}^{(k)}_4(x,t)\Big] ,\cr
M^{(k)}(x,t)_{2,2}&=&-M^{(k)}(x,t)_{1,1}.
\eeaa}\normalsize{}

Generically, the last recursion may create singularities at the zeros of $\det A^{(0)}(x,t)$, at $x= \infty$ and at singularities of the matrices $\left(A^{(k)}(x,t)\right)_{k\geq 0}$. In our present case, we know that we have
 $$
 A(x,\mathbf{t},\hbar) = \frac{\hat{A}(x,\mathbf{t},\hbar) }{ \beta(x)}
$$
where $\beta(x) = 1$ or $\beta(x) = x-a_i$ and $ \hat{A}(x,\mathbf{t},\hbar)$ is a $\mathfrak{sl}_2$ valued polynomial in $x$ of degree at most $1$. Thus, since $\det A^{(0)}(x,t)=\frac{1}{\beta(x)^2}\det \hat{A}^{(0)}(x,t)$, we may reduce the previous identities to
\bea \label{FullMRecursionbis}
M^{(k)}(x,t)_{1,1}&=&\frac{1}{\det \hat{A}^{(0)}(x,t)}\Big[\frac{1}{2} \hat{A}^{(0)}(x,t)_{1,1}\widehat{\text{RHS}}^{(k)}_1(x,t)+\frac{1}{2} \hat{A}^{(0)}(x,t)_{2,1}\widehat{\text{RHS}}^{(k)}_2(x,t)\cr
&&- \hat{A}^{(0)}(x,t)_{1,1}\sqrt{-\det \hat{A}^{(0)}(x,t)}\widehat{\text{RHS}}^{(k)}_4(x,t)\Big] ,\cr
M^{(k)}(x,t)_{1,2}&=&\frac{1}{\det  \hat{A}^{(0)}(x,t)}\Big[\frac{1}{2} \hat{A}^{(0)}(x,t)_{1,2}\widehat{\text{RHS}}^{(k)}_1(x,t)-\frac{1}{2} \hat{A}^{(0)}(x,t)_{1,1}\widehat{\text{RHS}}^{(k)}_2(x,t)\cr
&&- \hat{A}^{(0)}(x,t)_{1,2}\sqrt{-\det \hat{A}^{(0)}(x,t)}\widehat{\text{RHS}}^{(k)}_4(x,t)\Big] ,\cr
M^{(k)}(x,t)_{2,1}&=&\frac{1}{\det  \hat{A}^{(0)}(x,t)}\Big[-\frac{1}{2} \hat{A}^{(0)}(x,t)_{2,1}\widehat{\text{RHS}}^{(k)}_1(x,t)+\frac{1}{2} \hat{A}^{(0)}(x,t)_{1,1}\widehat{\text{RHS}}^{(k)}_3(x,t)\cr
&&- \hat{A}^{(0)}(x,t)_{2,1}\sqrt{-\det \hat{A}^{(0)}(x,t)}\widehat{\text{RHS}}^{(k)}_4(x,t)\Big] ,\cr
M^{(k)}(x,t)_{2,2}&=&-M^{(k)}(x,t)_{1,1}
\eea
with
\beq \label{hatRHS} 
\begin{pmatrix}\widehat{\text{RHS}}^{(k)}_1\\ \widehat{\text{RHS}}^{(k)}_2\\ \widehat{\text{RHS}}^{(k)}_3\\ \widehat{\text{RHS}}^{(k)}_4\end{pmatrix}\overset{\text{def}}{=}\begin{pmatrix} \beta(x)\partial_{t} M^{(k-1)}(x,t)_{1,1}-\underset{j=0}{\overset{k-1}{\sum}}\left[\hat{A}^{(k-j)}(x,t),M^{(j)}(x,t)\right]_{1,1}\\ 
\beta(x)\partial_t M^{(k-1)}(x,t)_{1,2}-\underset{j=0}{\overset{k-1}{\sum}} \left[\hat{A}^{(k-j)}(x,t),M^{(j)}(x,t)\right]_{1,2}\\
\beta(x)\partial_t M^{(k-1)}(x,t)_{2,1}-\underset{j=0}{\overset{k-1}{\sum}} \left[\hat{A}^{(k-j)}(x,t),M^{(j)}(x,t)\right]_{2,1}\\
\underset{j=1}{\overset{k-1}{\sum}}\left( M^{(j)}(x,t)_{1,1}M^{(k-j)}(x,t)_{1,1}+M^{(j)}(x,t)_{1,2}M^{(k-j)}(x,t)_{2,1}\right)\end{pmatrix}.
\eeq
Since the quantity $\beta(x)$ is no longer involved in denominators of equations \eqref{FullMRecursionbis} and \eqref{hatRHS} we get that the induction does not create any singularity at the zeros of $\beta(x)$ and thus that the only possible singularities of $M^{(k)}(x,t)$ are at the branchpoints (i.e. at the zeros of $\det A^{(0)}(x,t)$) or at $x=\infty$). Note that the conclusion remains valid if $\beta(x)$ is any arbitrary polynomial in $x$ (and not necessarily of degree at most $1$) since it simplifies in the computations. However, in this case, the control of the singularity at $x=\infty$ presented below may require some improvements.

\subsubsection{Study at $x=\infty$}
In this section we prove that the correlation functions are regular at $x=\infty$. As we will see below, the proof differs depending on the order of $\det  \hat{A}^{(0)}(x,t)$ at $x\to \infty$. 
\newline

 $\bullet$ Let us first discuss the case when $\det \hat{A}^{(0)}(x,t)\overset{x\to \infty}{=} \alpha x^2+O(x)$ with $\alpha\neq 0$. In other words $\det  \hat{A}^{(0)}(x,t)$ is a polynomial of degree $2$ in $x$. Since entries $\left( \hat{A}^{(0)}(x)\right)_{1\leq i,j\leq 2}$ are at most linear in $x$, formula \eqref{M0} immediately shows that $M^{(0)}(x)\overset{x\to \infty}{=}O(1)$ and therefore is regular at $x=\infty$. Then, an easy induction using \eqref{FullMRecursionbis} and the fact that $\beta(x)$ is at most linear, proves that for all $k\geq 0$: $M^{(k)}(x,t)\overset{x\to \infty}{=}O(1)$ and thus that the matrices $\left(M^{(k)}\right)_{k\geq 0}$ are regular at $x=\infty$. Indeed, by induction all numerators involved in \eqref{FullMRecursionbis} are at most quadratic in $x$ and thus dividing by $\det  \hat{A}^{(0)}(x,t)$ leads to at most constant terms at infinity. Eventually, using the definition of the correlation functions \eqref{DefWn}, we conclude that the correlation functions are regular at $x=\infty$.
\newline

$\bullet$ Let us now discuss the case when $\det  \hat{A}^{(0)}(x,t)\overset{x\to \infty}{=}\alpha x+O(1)$ with $\alpha\neq 0$. Note that section \ref{SectionLinearAuxMat} implies that this case may only happen for $\nu=0$, $r_0\geq 2$ and if $L_{0,r_0}$ does not have full rank. In the $\mathfrak{sl}_2$ case, this is equivalent to say that $L_{0,r_0}$ is of rank $1$. Using the $SL_2$ action presented in section \ref{SL2Action}, we may always get to the situation $L_{0,r_0}=\sigma_+$ and $L_{0,r_0-1}(t,\hbar)=C+l(t,\hbar)L_{0,r_0}$ where $C$ is a constant matrix (in the sense independent of $x$, $t$ and $\hbar$) . A typical example of this situation is the Painlev\'{e} I Lax pair (Cf. section \ref{P1Section}). Formula \eqref{ACase3} shows that we must have $A^{(0)}(x,t,\hbar)=L_{0,r_0}x+L_{0,r_0-1}(t,\hbar)+\beta_2(t,\hbar) A_{2,0,r_0-1}$. Moreover, since $A_{2,0,r_0-1} = 2 \left[x^{-r_0} L(x)\right]_+ =L_{0,r_0} $, we end up with 
\beq \label{A0SpecialCase} 
A(x,t,\hbar)=L_{0,r_0}x+C+\left(l(t,\hbar)+\beta_2(t,\hbar)\right)L_{0,r_0}\,\,\text{ with }\,\, L_{0,r_0}=\sigma_+ .
\eeq
Formula \eqref{M0} then immediately shows that we have
$$
M^{(0)}(x,t)=L_{0,r_0}x^{\frac{1}{2}}+\frac{1}{2}I_2+O\left(x^{-\frac{1}{2}}\right).
$$
We may now prove by induction that $M^{(k)}(x,t)\overset{x\to\infty}{=}O\left(x^{-\frac{1}{2}}\right)$ for $k\geq 1$. Indeed for $k\geq 1$ we have:
\bea \label{IntermediateOrders} \beta(x)\partial_t M^{(k-1)}(x,t)&=&O\left(x^{\frac{1}{2}}\right) \text{ since } \beta(x)\overset{x\to \infty}{=}O(x)\cr
\left[A^{(k)},M^{(0)}\right]&=&O\left(x^{\frac{1}{2}}\right)\,\,,\,\, \forall\, k\geq 1\cr
\left[A^{(k-j)},M^{(j)}\right]&=&O\left(x^{\frac{1}{2}}\right)\,\,,\,\, \forall\, k\geq 1\,,\,1\leq j\leq k-1\cr
\widehat{\text{RHS}}^{(k)}_{i}&=&O\left(x^{\frac{1}{2}}\right)\,\,,\,\, \forall\, i\in\{1,2,3\}\cr
\widehat{\text{RHS}}^{(k)}_4&=&O\left(x^{-1}\right)
\eea
Note that the first identity is valid for $k=1$ since $\partial_t M^{(0)}(x,t)=O\left(x^{-\frac{1}{2}}\right)$. The second and third identity hold since \eqref{A0SpecialCase} gives
\beqq 
A^{(k)}(x,t)=\left(l(t,\hbar)+\beta_2(t,\hbar)\right)^{(k)}L_{0,r_0}\propto L_{0,r_0},
\eeqq
so that the first two terms $L_{0,r_0}x^{\frac{1}{2}}$ and $\frac{1}{2}I_2$ do not contribute to the commutator $\left[A^{(k)},M^{(0)}\right]$. The fourth identity is a direct consequence of the first two using definition \eqref{hatRHS} while the fifth one is obvious from the induction assumption and the fact that $M^{(0)}$ is not involved in the sum defining $\widehat{\text{RHS}}^{(k)}_4$.
Eventually, inserting \eqref{IntermediateOrders} in \eqref{FullMRecursionbis} and the fact that $\hat{A}^{(0)}(x,t)=O(x)$ and $\det \hat{A}^{(0)}(x,t)=\alpha x+O(1)$ with $\alpha\neq 0$ shows that $M^{(k)}=O\left(x^{-\frac{1}{2}}\right)$ thus concluding the induction. 
Finally, the definition of the correlation functions \eqref{DefWn} (involving denominators $\frac{1}{x_i-x_{\sigma(i)}}$) shows that the correlation functions are regular at infinity (even if $M^{(0)}(x,t)=O\left(x^{\frac{1}{2}}\right)$ is not).

\subsection{Proof of property 6 : Identification of the two point function}
In order to prove the reconstruction by the topological recursion, we need to verify that $\omega_{0,1}$ and $\omega_{0,2}$ defined by the determinantal formulas identify with the data of the classical spectral curve. Using \eqref{DefWn}, this is equivalent to checking that
\beaa
 -\Tr( L^{(0)}(x,\mathbf{t})M^{(0)}(x,\mathbf{t})) &=&y(x)\cr
\frac{\Tr(M^{(0)}(x(z_1),\mathbf{t})M^{(0)}(x(z_2),\mathbf{t})) }{(x(z_1)-x(z_2))^2}x'(z_1)x'(z_2)&=&\frac{1}{(z_1-z_2)^2} .\cr
&&\eeaa
The first equation is automatically satisfied since $L^{(0)}(x,t)=\frac{A^{(0)}(x,t)}{\alpha(x,\mathbf{t})}$, $M^{(0)}=\frac{1}{2}I_2+\frac{1}{2\sqrt{-\det A^{(0)}}} A^{(0)}$ (Cf. \eqref{M0}) and the fact that $y=-\frac{\sqrt{\det A^{(0)}}}{\alpha(x,\mathbf{t})}$ as well as the fact that for $B\in \mathfrak{sl}_2$: $\Tr \left(B^2\right)=-2\det B$. 
\newline

$\bullet$ Let us verify that the second equation holds when $\hat{A}^{(0)}(x,t)=A_1(t)x+A_0(t)$ with $\det A_1\neq 0$. In this case we first observe that
$$
M^{(0)}(x,t)=\frac{1}{2}I_2+\frac{1}{2\sqrt{-\det \hat{A}^{(0)}(x,t)}} \hat{A}^{(0)}(x,t).
$$
Thus, we get $\det \hat{A}^{(0)}(x,t)=(\det A_1) x^2-\Tr(A_0A_1)x+\det A_0=(\det A_1) (x-a)(x-b)$ where the two branchpoints satisfy
$$
ab=\frac{\det A_0}{\det A_1} \text{ and } a+b=\frac{\Tr(A_0A_1)}{\det A_1}.
$$
Thus, we have
\bea \label{TrTr}\Tr(M^{(0)}(x(z_1),\mathbf{t})M^{(0)}(x(z_2),\mathbf{t}))&=&\frac{1}{2}+\frac{-2\det A_1 x(z_1)x(z_2)+\Tr(A_0A_1)(x(z_1)+x(z_2))+\det A_0}{4\sqrt{-\det A^{(0)}(x(z_1))}\sqrt{-\det A^{(0)}(x(z_1))}}\cr
&=& \frac{1}{2}+\frac{\det A_1\left( -2 x(z_1)x(z_2)+(a+b)(x(z_1)+x(z_2))+ab\right)}{4i^2\sqrt{\det A^{(0)}(x(z_1))}\sqrt{\det A^{(0)}(x(z_1))}}\cr
&=& \frac{1}{2}-\frac{ -2 x(z_1)x(z_2)+(a+b)(x(z_1)+x(z_2))+ab}{4\sqrt{(x(z_1)-a)(x(z_1)-b)}\sqrt{(x(z_2)-a)(x(z_2)-b)}} .\cr
&&
\eea
Using the parametrization $x(z)=\frac{a+b}{2}+\frac{b-a}{4}\left(z+\frac{1}{z}\right)$ giving $x'(z)=\frac{b-a}{4}\left(1-\frac{1}{z^2}\right)$ and
\beqq 
\sqrt{(x(z)-a)(x(z)-b)}=\frac{(b-a)}{4}\left(z-\frac{1}{z}\right),
\eeqq
a straightforward computation from \eqref{TrTr} gives
\beqq 
\frac{\Tr(M^{(0)}(x(z_1),\mathbf{t})M^{(0)}(x(z_2),\mathbf{t})) }{(x(z_1)-x(z_2))^2}x'(z_1)x'(z_2)=\frac{1}{(z_1-z_2)^2}.
\eeqq
\newline

$\bullet$ Let us verify that the second equation holds when $\hat{A}^{(0)}(x,t)=A_1(t)x+A_0(t)$ with $\det A_1=0$. We get $\det \hat{A}^{(0)}(x,t)=-\Tr(A_0A_1)x+\det A_0=-\Tr(A_0A_1)(x-a)$ where the only branchpoint satisfies $a=\frac{\det A_0}{\Tr(A_0A_1)}$. Note that $\Tr(A_0A_1)\neq 0$, otherwise the classical spectral curve would be trivial (no branchpoints). We have
\beaa
\Tr(M^{(0)}(x(z_1),\mathbf{t})M^{(0)}(x(z_2),\mathbf{t}))&=&\frac{1}{2}+\frac{\Tr(A_0A_1)(x(z_1)+x(z_2))-2\det A_0}{4\sqrt{-\det A^{(0)}(x(z_1))}\sqrt{-\det A^{(0)}(x(z_1))}}\cr
&=& \frac{1}{2}+\frac{\Tr(A_0A_1)\left(x(z_1)+x(z_2)-2a\right)}{4\sqrt{-\det A^{(0)}(x(z_1))}\sqrt{-\det A^{(0)}(x(z_1))}}\cr
&=& \frac{1}{2}+\frac{ x(z_1)+x(z_2)-2a}{4\sqrt{x(z_1)-a}\sqrt{x(z_2)-a}}.\cr
&&
\eeaa
Using the parametrization $x(z)=a+z^2$ giving $x'(z)=2z$ and $\sqrt{x(z)-a}=z$, we end up with
\beqq 
\Tr(M^{(0)}(x(z_1),\mathbf{t})M^{(0)}(x(z_2),\mathbf{t}))= \frac{1}{2}+\frac{ z_1^2+z_2^2}{4z_1z_2}=\frac{ (z_1+z_2)^2}{4z_1z_2}
\eeqq
so that
\beqq
 \frac{\Tr(M^{(0)}(x(z_1),\mathbf{t})M^{(0)}(x(z_2),\mathbf{t})) }{(x(z_1)-x(z_2))^2}x'(z_1)x'(z_2)=\frac{ (z_1+z_2)^2}{z_1^2-z_2^2}=\frac{1}{(z_1-z_2)^2} .
\eeqq

 We conclude that in both situations (one or two branchpoints), the existence of a linear auxiliary matrix presented in section \ref{SectionLinearAuxMat} is sufficient to prove the identification of the initial data of the topological recursion (classical spectral curve and Bergman kernel) with their corresponding quantities given by the determinantal formulas.
 
\subsection{Proof of property 5 : Leading order in $\hbar$}
This follows from the other properties using the double recursion introduced in \cite{MarchalIwaki}.

\end{document}